\documentclass[10pt, a4paper]{article}
\usepackage[utf8]{inputenc}
\usepackage[english]{babel}
\usepackage{ragged2e}

\usepackage{geometry}
\usepackage{setspace}            

\vspace{3mm} 
\usepackage{graphicx, tabularx}
\usepackage{adjustbox}
\usepackage{caption}            
\usepackage{rotating}           

\usepackage{tabularx}
\usepackage{subcaption}
\newcommand\fnote[1]{\captionsetup{font=small}\caption*{#1}}

\usepackage{booktabs}
\usepackage{multirow}
\usepackage{threeparttable}
\usepackage[online]{threeparttablex}
\usepackage{float}
\usepackage[bottom]{footmisc}

\usepackage{relsize}
\DeclareUnicodeCharacter{2212}{-}

\usepackage{dcolumn}    
    \newcolumntype{d}[1]{D{.}{.}{#1}}

\usepackage{natbib}
\usepackage[colorlinks=true, allcolors=blue, backref=page]{hyperref}

\usepackage{amssymb, amsfonts, amsmath}
\usepackage{bm}

\usepackage{enumerate, enumitem}

\usepackage{pdflscape}

\usepackage{longtable}
\usepackage{xtab,afterpage}
 
\usepackage{comment}

\title{Going Green: Estimating the Potential of Green Jobs in Argentina \thanks{The usual disclaimer applies. Corresponding author: \href{mailto:delavegapc@gmail.com}{delavegapc@gmail.com}}}

\author{
    Natalia Porto\thanks{Instituto de Investigaciones Económicas, Facultad de Ciencias Económicas, Universidad Nacional de La Plata, Argentina.}
    \and
    Pablo de la Vega\thanks{Instituto de Investigaciones Económicas, Facultad de Ciencias Económicas, Universidad Nacional de La Plata, Argentina}
    \and
    Manuela Cerimelo\thanks{Instituto de Investigaciones Económicas, Facultad de Ciencias Económicas, Universidad Nacional de La Plata, Argentina}
}

\date{Last updated: \today}

\begin{document}

\maketitle

\begin{abstract}
\noindent This paper aims to identify and characterize the potential of green jobs in Argentina, i.e., those that would benefit from a transition to a green economy, using occupational green potential scores calculated in US O*NET data. We apply the greenness scores to Argentine household survey data and estimate that 25\% of workers are in green jobs, i.e., have a high green potential. However, when taking into account the informality dimension, we find that 15\% of workers and 12\% of wage earners are in formal green jobs. We then analyze the relationship between the greenness scores (with emphasis on the nexus with decent work) and various labor and demographic variables at the individual level. We find that for the full sample of workers the green potential is relatively greater for men, the elderly, those with very high qualifications, those in formal positions, and those in specific sectors such as construction, transportation, mining, and industry. These are the groups that are likely to be the most benefited by the greening of the Argentine economy. When we restrict the sample to wage earners, the green potential score is positively associated with informality.
\end{abstract}



\newpage


\section{Introduction}

Human-induced climate change is probably the most serious problem that humankind is facing \citep{IPCC2021, IPCC2022}. Because of the threat to climate stability, biodiversity, and economic development, there has been a great deal of academic and political interest in designing policies towards green transition and sustainable development. According to \cite{Dierdorff2009}, a ‘green economy’ involves those economic activities that focus specifically on reducing pollution. For instance, activities that directly or indirectly use less fossil fuels or reduce the greenhouse gas emissions, as well as activities that promote energy efficiency or recycling materials, would be embodied in the green economy. In addition to reducing environmental risks, the United Nations Environment Programme adds that a ‘green economy’ should also have a positive impact on welfare and social equity.\footnote{For more information see \href{https://www.unep.org/regions/asia-and-pacific/regional-initiatives/supporting-resource-efficiency/green-economy}{here}.}. Similarly, the International Labour Organization’s (ILO)’s Green Jobs initiative highlights that the creation of decent work in the new green activities and the implementation of social protection policies to mitigate the effects on the sectors that need transforming are key to ensuring this inclusiveness of the transition to greener economies\footnote{For more information see \href{https://tinyurl.com/greenjobs-iniciative}{here}.}.  

There is broad consensus that the "green transition" will generate both environmental benefits and new opportunities for economic development in the long run \citep{Consoli2016}. For instance, \cite{Wallach2022} predicts that the transition to greener forms of energy will create more than 10 million new jobs worldwide by 2030, a figure that exceeds the number of jobs expected to be lost in most polluting sectors, such as fossil fuels. In fact, the shift to hiring more green jobs is already underway worldwide. As noted by \cite{Kimbrough2021}, as a result of governments and companies committing to their climate and sustainability goals, the demand for green skills has been steadily increasing since 2017. For Latin America and the Caribbean, \cite{Saget2020} argue that cleaner forms of production, such as decarbonization, have the potential to generate 15 million net jobs by 2030. 

Acknowledging the above, several countries have been developing strategies to foster the implementation of green technologies, the creation of sustainable industries, the reduction in current pollution levels, among other goals\footnote{See, for example, initiatives in advanced economies \href{https://ec.europa.eu/reform-support/what-we-do/green-transition_en}{here} and \href{https://tinyurl.com/oecd-greengrowth}{here}, and \href{https://www.argentina.gob.ar/produccion/cep/investigaciones-sobre-la-estructura-productiva/ambiente}{here} from a perspective of the development and diffusion of technologies.}. In particular, Argentina has been making efforts to implement environmentally friendly policies. Since 2015, the country has launched several programs that seek to promote renewable energy sources (like RenovAr and Probiomasa programs) and has included goals of sustainable transition to greener economies in the productive, industrial and technological development plan for 2030 (in Spanish, Plan Argentina Productiva 2030)\footnote{Specifically, this national strategy intends to develop green industries, adapt the existing ones to reduce their environmental impact, and create more than 2 million green jobs outside the metropolitan area. For more information see \href{https://www.argentina.gob.ar/produccion/argentina-productiva-2030}{here}.}. Additionally, since Argentina is part of the United Nations Framework Convention on Climate Change, it has established several National Determined Contributions which intend to contribute to the goals of the Paris Agreement\footnote{Primarily, with the goal of ‘holding the global average temperature increase to well below 2°C in comparison to pre-industrial levels, and pursuing efforts to limit that temperature increase to 1.5°C in relation to pre-industrial levels’.}.   

Despite the consensus over the expected long-term benefit of the green transition, in many countries it would take time to transform their economies into greener ones because changes in the production methods and in the workforce skills could be necessary. As a result, in the short or medium run, the cost of adopting greener policies or activities could increase considerably for some sectors, and frictions in the labor market may emerge, resulting in winners and losers in terms of employment, as this transition would probably change the allocation of workers across occupations and sectors. Like in many other labor market transformations, the burden costs will be higher for those workers who perform tasks and use skills that are less compatible with the new labor market conditions. Moreover, accelerating human capital accumulation is crucial, considering that the transition to greener activities will be almost straightforward for those workers with higher skills \citep{IMF2022}. 

In consequence, it is crucial to identify which jobs are more/less compatible with the conditions imposed by the greening of an economy in order to manage the transition efficiently. When addressing the labor market implications of the green transition, \cite{Dierdorff2009} suggest that it is convenient to adopt an occupational approach and thus define green jobs as those affected by green activities and technologies. In particular, these green economy activities or technologies can increase the demand for existing occupations, shape the work and worker requirements needed for occupational performance, or generate new work and worker requirements. The extent to which the green activities can alter the labor market conditions is what \cite{Dierdorff2009} denote as the ‘greening of occupations’. It is worth noting that non-green occupations are not necessarily ‘dirty’ or ‘brown’ but they are not affected by the greening of an economy. Indeed, after defining the concepts of 'green economy', 'green occupation', and 'the greening of occupations', \cite{Dierdorff2009} review the literature and list 12 major green economy sectors where green occupations can be found (renewable energy generation, energy efficiency and trading, agriculture and forestry, recycling and waste management, transportation, green construction, and manufacturing, among others). In that way, green occupations can be found in sectors that may be classified as 'dirty' based on, for example, definitions which use sectoral emissions.

In light of the above, this paper aims to characterize the potential of green jobs in Argentina -defined as the ability of an occupation to be able to perform green tasks \citep{Lobsiger2021}. We identify which type of workers would win or lose as a consequence of the transition to a green economy, using occupational green potential scores calculated with US O*NET data. While most of previous analyses focused on developed economies, our paper is the first to apply an occupation-based approach with a greenness perspective in a developing country like Argentina. In addition, we study the interrelation between the potential of green jobs and the ‘decent work’ dimension. Specifically, we analyze how labor informality, as a proxy for decent work, affects the effective green transition of occupations. This dimension becomes crucial in countries like Argentina with a high share of informal employment.

Our results show that 25\% of workers are in green jobs, i.e., have a high green potential. However, when considering the informality dimension, we find that 15\% of workers and 12\% of wage earners are in formal green jobs. We also find that the green potential is relatively greater for men, the elderly, those with very high qualifications, and in specific sectors such as construction, transportation, mining, and industry. These are the groups that are likely to be the most benefited by the greening of the Argentine economy. Moreover, when considering the informality dimension, we show that for the full sample of workers there is a negative relation with all the green potential measures. However, when we restrict the sample to wage earners, we find that the opposite holds true: informality is positively associated with both the binary definition of green potential and the high green potential measure. Therefore, those workers in formal jobs are also likely to benefit from the green transition, although for wage earners this transition may be incompatible with decent work.

The remainder of the paper is organized as follows. Section \ref{sec:lit_rev} describes the literature review. Section \ref{sec:data} presents the sources of information used. Section \ref{sec:emp_ev} deals with descriptive analyses. Finally, Section \ref{sec:results} presents the results and Section \ref{sec:conclusion} concludes.

\section{Literature review} \label{sec:lit_rev}

This section provides an overview of the relevant literature that seeks to measure the green potential of jobs\footnote{Another strand of the literature has focused on evaluating the effects of environmental policies, which are supposed to ease the transition to greener economies, on various macroeconomic outcomes. See, for example, \cite{Hafstead2018} for a more comprehensive discussion on the effects of these kinds of policies using computable general equilibrium models.}. Given the likely heterogeneous effects of the green transition on economies and on their labor markets in the short run, it is highly relevant to learn about which types of jobs and skills may be affected by this transformation. According to the ILO Green Jobs Initiative, ‘green jobs’ refer to those that take place in economic activities that are more environmentally sustainable than the conventional alternative and which also offer working conditions that meet accepted standards of ‘decent work’ \citep{Jarvis2011}. The concept of ‘decent work’ refers to the necessary conditions of an employment relationship to be carried out ‘in conditions of freedom, equity, security and human dignity’. For work to be considered decent, workers must have: (1) productive jobs with a fair wage, (2) good working conditions, (3) social protection, (4) labor rights, (5) equal opportunities between genders and (6) a say in decisions which will affect their lives \citep{ILO2013}\footnote{\cite{Eurofound2012} develops an alternative approach to decent work that consists in measuring job quality based on extrinsic job features – ‘earnings’ and ‘prospects’-, alongside a larger set of intrinsic characteristics of the job itself - ‘intrinsic job quality’ (work and its environment) and ‘working time quality’. Contrary to the ILO framework of decent work, \cite{Eurofound2012} made explicit choices regarding to the main aspects to prioritize and the set of indicators (mostly objective) to consider for each dimension, based on their impact on workers’ well-being \citep{Cazes2015}.}. Green jobs are thus both environmentally sound and ‘decent’ in social terms. However, the existing literature has not yet agreed on a widely accepted definition of what a green job is, and most importantly, on how to identify these types of jobs in practice. Indeed, the empirical literature has taken three main approaches to identify green jobs based on: i) the industry affiliation; ii) the production methods used; and iii) the task content of occupation. Below, we briefly describe each approach and discuss the pros and cons. A more comprehensive explanation can be found, for example, in \cite{Martinez2010}, \cite{Consoli2016}, \cite{Bowen2018}, and \cite{Vona2021}.

According to the industry affiliation approach, those sectors that produce goods and services that contribute to the protection of the environment or the conservation of natural resources are green sectors, and all employees working in that sector are considered green workers \citep{Eurostat2019, Yi2014,Yi2015}. However, as pointed out by \cite{ILO2018} and \cite{Esposito2017} this definition considers neither the jobs that eventually improve production processes with respect to their environmental impact nor the skills necessary to carry out environmentally friendly activities to achieve a sustainable economy. Besides, this approach is unable to identify green (brown) jobs in brown (green) industries.
The second approach consists in selecting employees of green processes, that is, specialized in activities for the protection of the environment, such as active waste management, treatment, and recycling \citep{USCommerce2010}. Nevertheless, \cite{Vona2019} noted that although this method allows a sharp identification due to the specificity of these activities, it neglects activities devoted to the whole re-design of products that are often carried out by specialized suppliers of machinery and engineering and architecture solutions. 

Contrary to the approaches discussed above, the third method applies the task-based approach  \citep{autordorn2013, autor2013, acemoglu2011} identifying green jobs according to the number of green tasks that a given occupation requires the worker to do. In particular, this strategy addresses the problem discussed when using the industry approach: much of the variation in the share of green employment is observed within rather than between industries \citep{Vona2019}. The key issue with this approach is to find appropriate data for the determination of green jobs and skills\footnote{In fact, for countries like Argentina, it is also difficult to find appropriate/available data for identifying green jobs using either the industry or the production approach.}. Most of the studies exploiting this approach use the US Occupational Information Network (O*NET) data.  The O*NET database contains not only detailed information on the task and skill content of occupations but also detailed text descriptions for a subset of tasks specific to each occupation. Furthermore, the ‘Green Economy Program’ developed by O*NET details the work tasks of green jobs allowing researchers to understand the changes in occupation and skill requirements that may be triggered when a country transitions to a greener economy \citep{Elliott2021}. This information can then be used to identify green jobs based on two broad types of definitions: i) a binary definition where an occupation is considered either green or non-green; ii) a continuous definition of occupational greenness that exploits information on the greenness of the task content of occupations \citep{Vona2021}. 

Several studies have relied on the O*NET data to measure and characterize green employment in the US. \cite{Consoli2016} classify occupations as green or non-green and show that green occupations in the US require higher levels of cognitive and interpersonal skills, higher levels of education, greater work experience, and greater on-the-job training. A novel approach by \cite{Vona2018} computes a continuous greenness measure for each occupation. On this basis, they assess the importance of green skills in the US and find that green occupations have a higher technical skill requirement even though they note that the overall skills gap between green and brown occupations is relatively small. In addition, \cite{Vona2019} study the characteristics of green jobs between 2006 and 2014 and suggest that green occupations yield a wage premium in contrast to non-green jobs. While using O*NET’s broad definition of green jobs, \cite{Bowen2018} find that 19.4\% of the US workforce is part of the green economy, although most green employment is ‘indirectly’ rather than ‘directly’ green, with only 13\% of the total workforce using any specifically green tasks in their jobs.

Research on this topic using O*NET to target green jobs has also extended outside the US, where occupational level data availability is a limitation. \cite{Rutzer2020} predict the ‘green potential’ of International Standard Classification of Occupations (ISCO) occupations based on their corresponding skills using again O*NET information. They evaluate the distributional consequences of greening the economy in 19 European countries for the period 1992-2010, and find heterogeneous responses between countries and occupations. For instance, occupations with relatively high green potential, such as science or engineering, are expected to benefit from greening, while low green potential occupations, like elementary occupations or clerical support workers, could lose. Similarly, \cite{Lobsiger2021} measure the green potential of occupations in Switzerland and estimate that 16.7\% of employment is in occupations with high green potential. Moreover, these workers are, on average, younger, more often men, have a higher level of education and a higher probability of having immigrated than employed persons in occupations with low green potential. \cite{Valero2021} use occupational classifications from O*NET applied to individual-level survey data to quantify and describe green jobs in the UK and European Union. Their results reveal that the share of green jobs ranges from 17\% (Greece) to 22\% (Germany), and they suggest that these kinds of jobs tend to be held by older, male workers, with higher educational level, and with permanent contracts. The main criticism that these studies have received is that the task content varies depending on the level of development and, therefore, it would not be correct to extrapolate estimates based on the United States to other countries, particularly emerging ones \citep{Dicarlo2016, LoBello2019}. 

Although several studies have estimated and described the share of green jobs in the economy, most of them are based solely on developed countries. Green jobs assessments in the developing world are scarce, since these countries do not usually have as much available data as the developed ones. To our knowledge, few studies have focused on Latin America and even fewer on Argentina. For Latin America and the Caribbean, \cite{Saget2020} study the possible impacts of an emissions reduction strategy on labor markets. The authors show that by 2030, structural changes in energy and food production and consumption patterns can eventually result in 15 million more net jobs in the region compared to a business-as-usual scenario. At the same time, they underline that many workers will have to update their skills to meet the demand of emerging sectors, and many firms will have to adopt new technologies and adapt to new ways of doing business. As part of the ILO’s Green Jobs initiative, \cite{Ernst2019} estimate the potential of green jobs in Argentina by sector and company and find that between 4\% and 7\% of jobs were green in 2015. In particular, most green jobs were in manufacturing, transport, the agriculture, livestock, forestry and fisheries sector, and water supply and waste management.

\section{Data sources and variable definition} \label{sec:data}

To identify the potential of green jobs in Argentina we exploit various data sources. First, we rely on the Permanent Household Survey (PHS) microdata from the Instituto Nacional de Estadísticas y Censos (INDEC; in English, National Institute of Statistics and Censuses) of Argentina. The PHS consists of piled cross-sections that collect data from households with a continuous frequency and its geographic scope reaches 31 urban agglomerations (cities). In particular, to gain sample size, we use the PHS data from 2015q1 to 2021q3\footnote{After the presidential change in 2015, the publication of the PHS surveys from 2015q3 to 2016q1 were put under review and finally not published.}. The final sample is restricted to those between 15 and 65 years old to avoid the influence of educational and retirement decisions on labor market participation.

Second, since there is no information on the task composition of occupations for Argentina, we estimate occupational greenness scores based on US data and then extrapolate it to the Argentine occupational structure. Specifically, we calculate the greenness scores proposed in \cite{Vona2021} in the O*NET dataset at the Standard Occupational Classification (SOC) level, and then extrapolate them to the Argentine occupational classification to impute them to each person employed in the PHS. This procedure has already been applied in several papers \citep{Rutzer2020, Lobsiger2021, Elliott2021, Valero2021} and is common in the automation literature \citep{Gasparini2020, Brambilla2021} and more recently in the teleworking literature \citep{Albrieu2020, Bonavida2020, delavega2021}. As stated in the previous section, these studies have been criticized because the task content varies depending on the level of development and, therefore, it would not be correct to extrapolate estimates based on the United States to other countries, particularly emerging ones \citep{Dicarlo2016, LoBello2019}. However, we lack alternatives based on data availability. The rest of this section explains how the greenness scores are calculated.

As explained in \cite{Vona2021}, O*NET identifies three groups of green occupations that will be affected by the greening of an economy: (i) occupations that are expected to experience an increase in demand (Green Increased Demand); (ii) occupations that will see major changes to the tasks content of work (Green Enhanced Skills); and (iii) occupations that did not exist before and that will be created (Green New and Emerging)\footnote{It is worth remembering that non-green occupations are not necessarily ‘dirty’ or ‘brown’ but are not affected by the greening of an economy.}. For (ii) and (iii), O*NET also identifies green tasks within each occupation, but not for (i) because they may benefit only indirectly from the greening of an economy. In consequence, green jobs can be identified by using O*NET data in two ways: i) a binary definition where an occupation is considered either green or non-green; ii) a continuous definition of occupational greenness that exploits information on the greenness of the task content of occupations. Following \cite{Vona2021}, we calculate the greenness of an occupation $j$ as follows:
\begin{align}
greenness_{j}=\frac{\# \text { green tasks }}{\# \text { total tasks }}
\end{align}
which takes values greater than zero only for Green Enhanced Skills and Green New and Emerging occupations\footnote{Given that O*NET also provides data on the importance of each task within an occupation, a weighted version of this indicator can be calculated. However, according to \cite{Vona2021}, the correlation between the unweighted and the weighted version is extremely high, thus the use of such weights is unnecessary.}. As stated in \cite{Vona2021}, the greenness indicator can be considered as a proxy for the amount of time spent on green activities and technologies in the average job post within a certain occupation. Additionally, O*NET identifies core tasks within each occupation, thus we can also calculate a more restrictive score:
\begin{align}
greenness\_core_{j}=\frac{\# \text { green core tasks }}{\# \text { total core tasks }}
\end{align}

In summary, we have three greenness indicators: 1) a binary definition where an occupation is considered either green or non-green (\textit{green\_occ}); 2) a task-based indicator that considers the proportion of green tasks on total tasks within an occupation (\textit{greenness}); 3) a task-based indicator that considers the proportion of green core tasks on total core tasks within occupation (\textit{greenness\_core}).

Once we have the greenness indicators at the 8-digit SOC level, the goal is to transfer them to the 2-digit ISCO classification (International Standard Classification of Occupations; the classification used in the Argentine Household Surveys). A crosswalk between the 6-digit level SOC and 4-digit ISCO classifications is provided by the US Bureau of Labor Statistics. Therefore, the first step consists in calculating the simple average of the greenness indicators at the 6-digit SOC\footnote{This step is very common in the literature (see, for example, \cite{OECD2017}, \cite{Goos2014}, \cite{Consoli2016}, \cite{Vona2018}, \cite{Elliott2021}, \cite{Rutzer2020}).}. The most challenging procedure is the mapping of 6-digit SOC codes to 2-digit ISCO codes. Given the many-to-many mapping, taking a weighted average using the SOCs employment as weights would put disproportionate weight on those SOCs that map to a larger number of ISCOs\footnote{Indeed, other papers that extrapolate greenness indicators from O*NET to ISCO apply a simple average at each step. See, for example, \cite{Elliott2021} and \cite{Rutzer2020}. In addition to this method, \cite{Valero2021} use a different approach for the last step considering a UK occupation as green if at least one of its matched occupations from O*NET is green.}. In consequence, following \cite{Dingel2020}, when an SOC maps to multiple ISCOs, we allocate the SOC's employment weight across the ISCOs in proportion to the ISCOs' employment shares\footnote{Formally, for each SOC-ISCO combination, the resulting weight is as follows: \[ weight_{SOC, ISCO}=L_{SOC}^{BLS} \frac{L_{S O C, I S C O}^{I L O}}{\sum_{I S C O} L_{S O C, I S C O}^{I L O}} \] where $L_{SOC}^{BLS}$ is the 6-digit SOC employment from the US Bureau of Labor Statistics, and $L_{SOC,ISCO}^{ILO}$ is the 2-digit ISCO employment from the International Labour Organization (ILO).}. 

\begin{figure}[H]
    \centering
    \caption{Greenness over ISCO-2d occupations}
    \includegraphics[width=\textwidth]{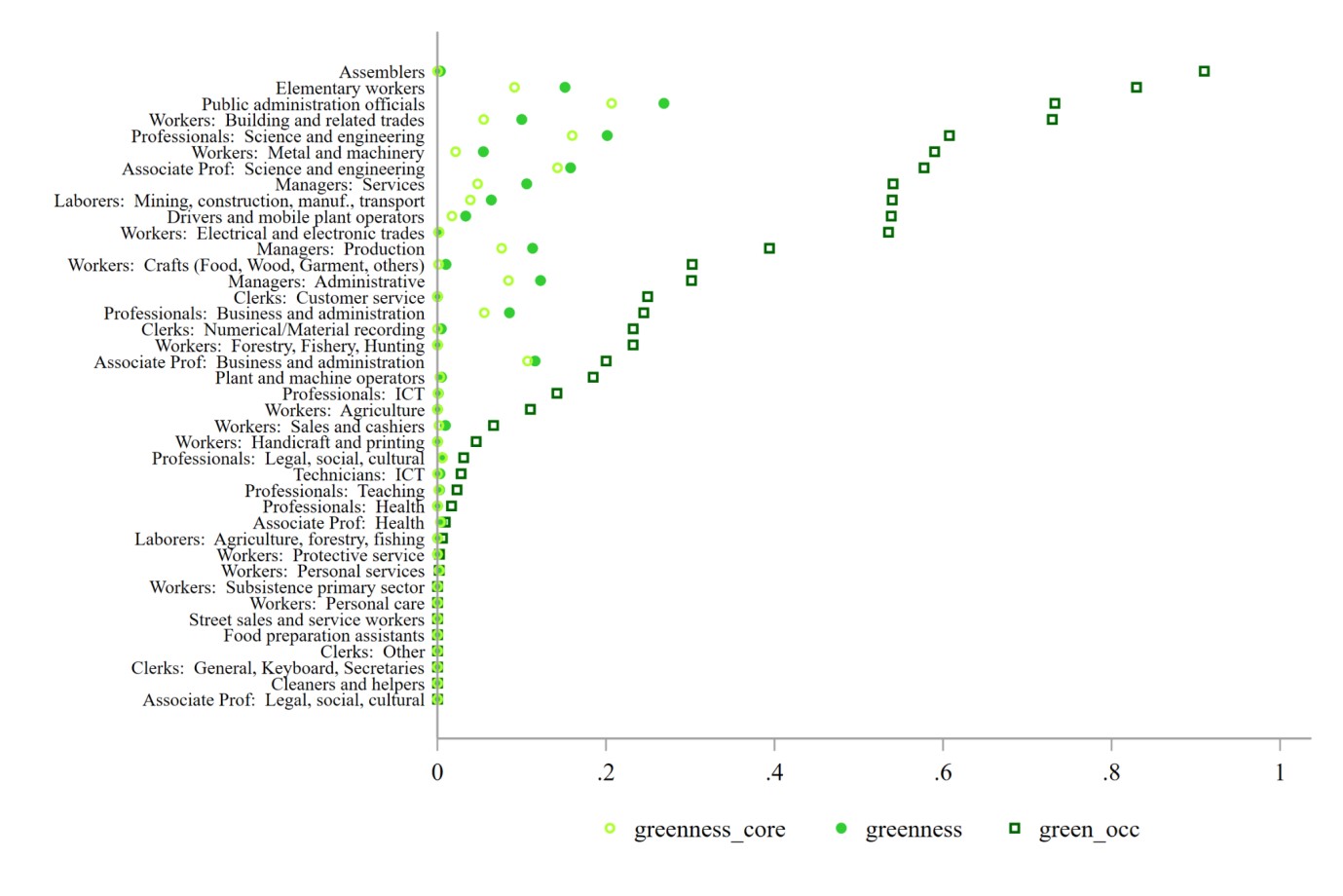}
     \fnote{{\smaller Own elaboration. The figure shows the estimated greenness scores at the 2-digit ISCO level. \textit{green\_occ} refers to a binary definition where an occupation is considered either green or non-green; greenness is a task-based indicator that accounts for the proportion of green tasks on total tasks within an occupation; and \textit{greenness\_core} is a task-based indicator that accounts for the proportion of green core tasks on total core tasks within an occupation.}}
         \label{fig:fig1}
\end{figure}

The resulting greenness scores at the 2-digit ISCO level are shown in Figure \ref{fig:fig1}. The results are very similar to previous literature \citep{Vona2018, Elliott2021, Rutzer2020, Lobsiger2021}. Occupations with the highest green potential are science and engineering professionals, managers, assemblers, whereas among those with the lowest green potential we find health and personal services workers, and clerks.

\begin{figure}[H]
    \centering
    \caption{Greenness over ISCO-2d occupations in PHS}
    \includegraphics[width=\textwidth]{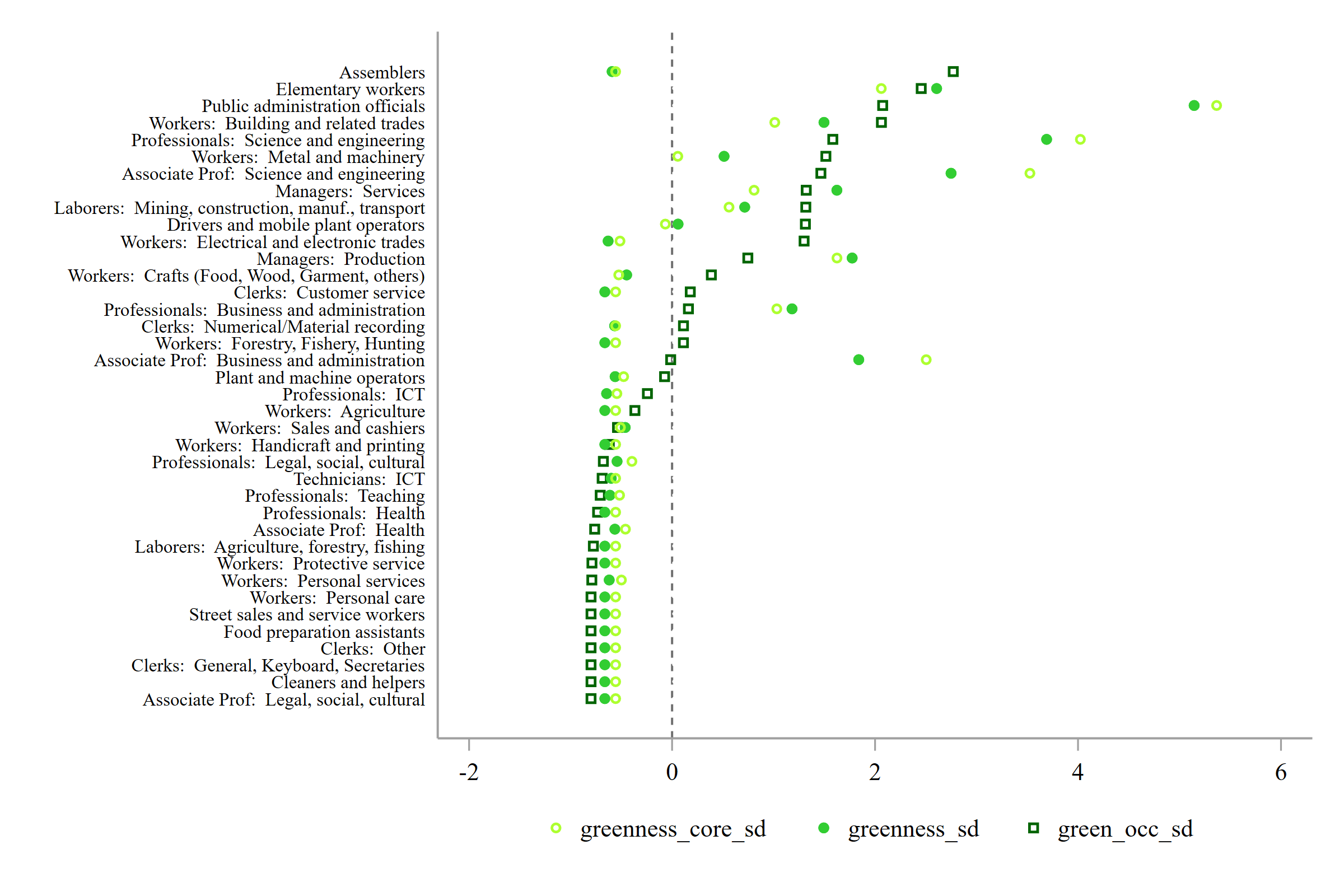}
     \fnote{{\smaller Own elaboration based on PHS. The figure shows the average greenness scores at the 2-digit ISCO level. \textit{green\_occ} refers to a binary definition where an occupation is considered either green or non-green; greenness is a task-based indicator that accounts for the proportion of green tasks on total tasks within an occupation; and \textit{greenness\_core} is a task-based indicator that accounts for the proportion of green core tasks on total core tasks within an occupation. The scores were standardized such that for the entire sample of workers, each greenness measure has a mean of zero and a standard deviation of one. Survey weights were used.}}
         \label{fig:fig2}
\end{figure}

Once we have the greenness scores at 2-digit ISCO, we impute them to each person employed in the Argentine PHS\footnote{It is worth noting that the greenness scores are fixed within the whole period, thus changes in the green potential of jobs must be understood as triggered by changes in the occupational structures.}. To ease comprehension, we perform a standardization procedure as in \citep{acemoglu2011} such that for the entire sample of workers, each greenness measure has a mean of zero and a standard deviation of one. The resulting greenness scores at the 2-digit ISCO level are shown in Figure \ref{fig:fig2}.

Finally, to identify green jobs we follow previous literature which defines those jobs as high green potential occupations. \cite{Elliott2021} consider an individual to be a green worker if their corresponding occupational greenness score is greater than the average greenness (0.189 in their sample). Similarly, \cite{Lobsiger2021} define high-green-potential occupations as those with green potential larger or equal to 0.5. This threshold was adopted because they find a significant positive association between an increase in the implicit emission tax and demand only for occupations with a green potential equal to or above 0.5. Moreover, the median green potential is 0.27 in their sample, thus high-green-potential occupations include occupations that have more than one and a half times the median green potential. 

We follow a similar approach defining high green potential occupations as those with greenness scores greater than a factor times the mean green potential. That factor is set to maximize the correlation between the sectoral shares to the ones obtained by \cite{Ernst2019}, who follow the methodology proposed by the ILO’s Green Jobs initiative to estimate green jobs in Argentina. According to the definition of ILO, green jobs are those that, by meeting decent work standards, contribute to the preservation and restoration of the environment. To compare our estimates with those of \cite{Ernst2019}, we additionally define formal green jobs as those workers with high green potential who are in formal positions\footnote{\cite{Ernst2019} do not consider unregistered green jobs as green jobs because of poor working conditions.}. As \cite{Ernst2019}, we consider formal workers as wage employment contributing to the social security system. Moreover, we only use 2015 data to compare estimates in the same dataset. 

\begin{figure}[H]
    \centering
    \caption{Figure 3. \cite{Ernst2019} vs Our Estimates in 2015}
    \includegraphics[width=\textwidth]{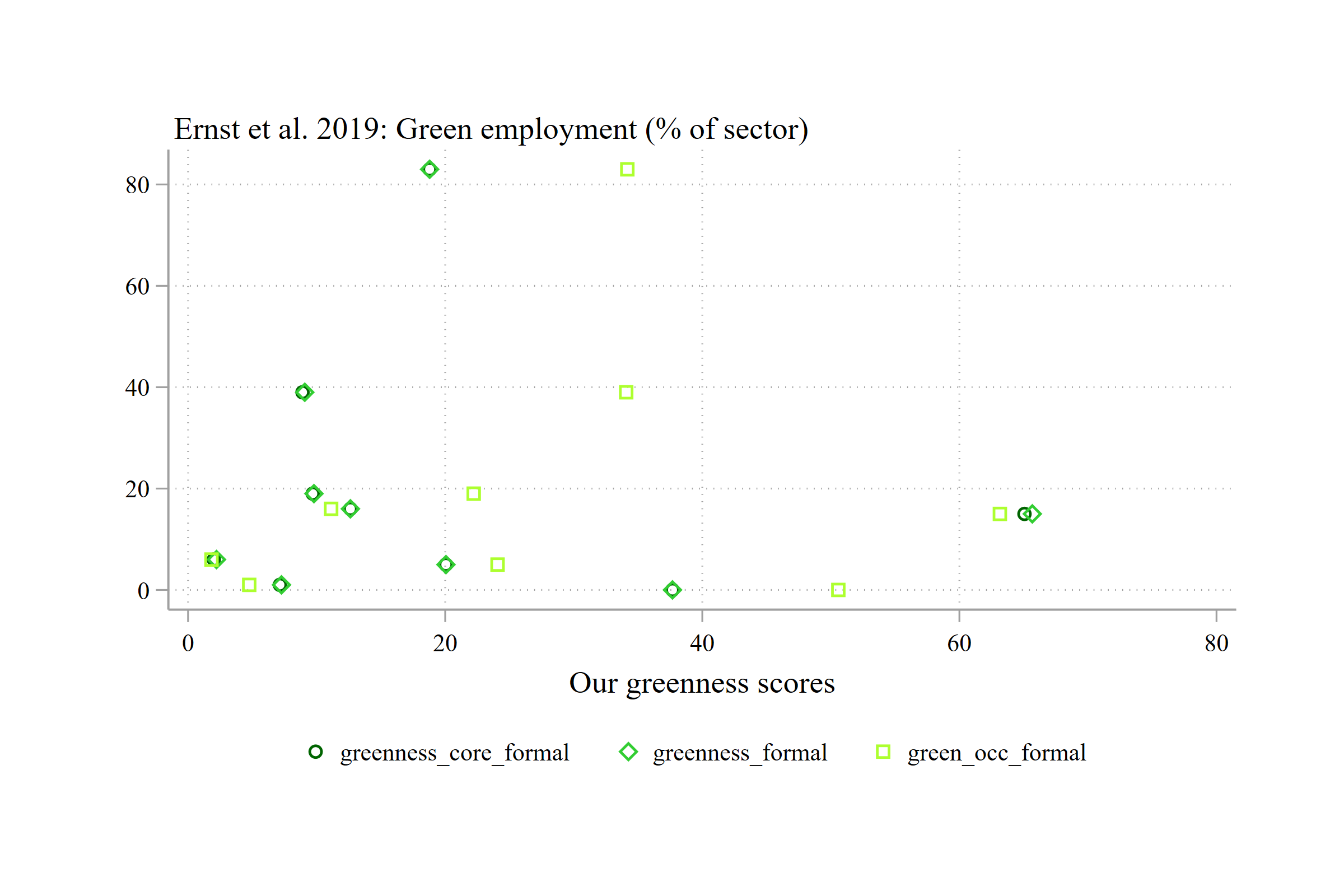}
     \fnote{{\smaller Own elaboration based on PHS. The figures compare our results with those in Table 1 of \cite{Ernst2019}. Calculations for wage earners. \textit{green\_occ} refers to a binary definition where an occupation is considered either green or non-green; greenness is a task-based indicator that accounts for the proportion of green tasks on total tasks within an occupation; and \textit{greenness\_core} is a task-based indicator that accounts for the proportion of green core tasks on total core tasks within an occupation. Survey weights were used. Each dot represents one of the following sectors: Agricultural and fishing; Construction; Electricity and Gas; Water; Food and accommodation; Industry; Mining; Transport and Communications; Rest.}}
         \label{fig:fig3}
\end{figure}

\begin{figure}[H]
    \centering
    \caption{High Green Potential Jobs, 2015q1-2021q3}
    \includegraphics[width=\textwidth]{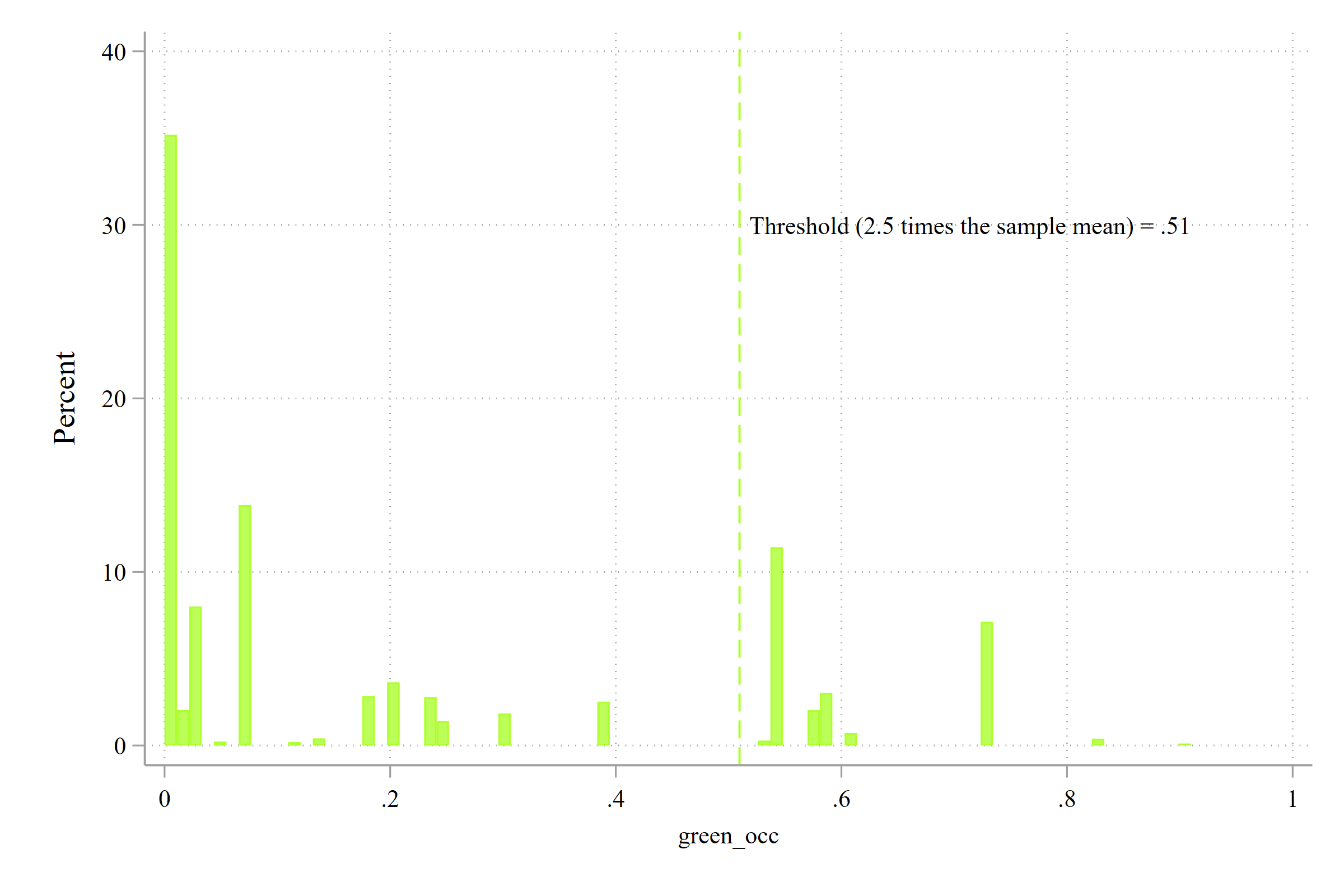}
     \fnote{{\smaller Own elaboration based on PHS. The figure shows the distribution of the green potential score. Every occupation with a score above 0.51 (2.5 times the sample mean) is classified as a green job (i.e., has a high green potential). Survey weights were used. Calculations for the period 2015q1-2021q3. \textit{green\_occ} refers to a binary definition where an occupation is considered either green or non-green.}}
         \label{fig:fig4}
\end{figure}

Figure \ref{fig:fig3} compares our estimates with those of \cite{Ernst2019}. We find that, in 2015, between 9\% and 12\% of wage earners were in formal green jobs, which is somewhat higher than the 7\% that \cite{Ernst2019} find. The correlation between the sectoral shares of green jobs is maximized when using the \textit{green\_occ} scores and a multiplying factor of 2.5. In consequence, we will use this greenness potential measurement in the rest of the paper. However, it is worth remembering that the correlation between the different greenness indicators is above 80 percent, so the results will vary little by using one or another. The whole distribution of the \textit{green\_occ} scores for the complete period under analysis is shown in Figure \ref{fig:fig4}, along with the said threshold separating green and non-green jobs. Besides this quantitative approach to the choice of the greenness measure for the rest of the paper, using the \textit{green\_occ} score has the advantage of also considering as green not only those that would experience work requirement changes or that will be newly created, but also those jobs that will experience a demand increase due to the green transition.

\section{Descriptive analysis} \label{sec:emp_ev}

\begin{figure}[H]
    \centering
    \caption{Greenness over age}
    \includegraphics[width=\textwidth]{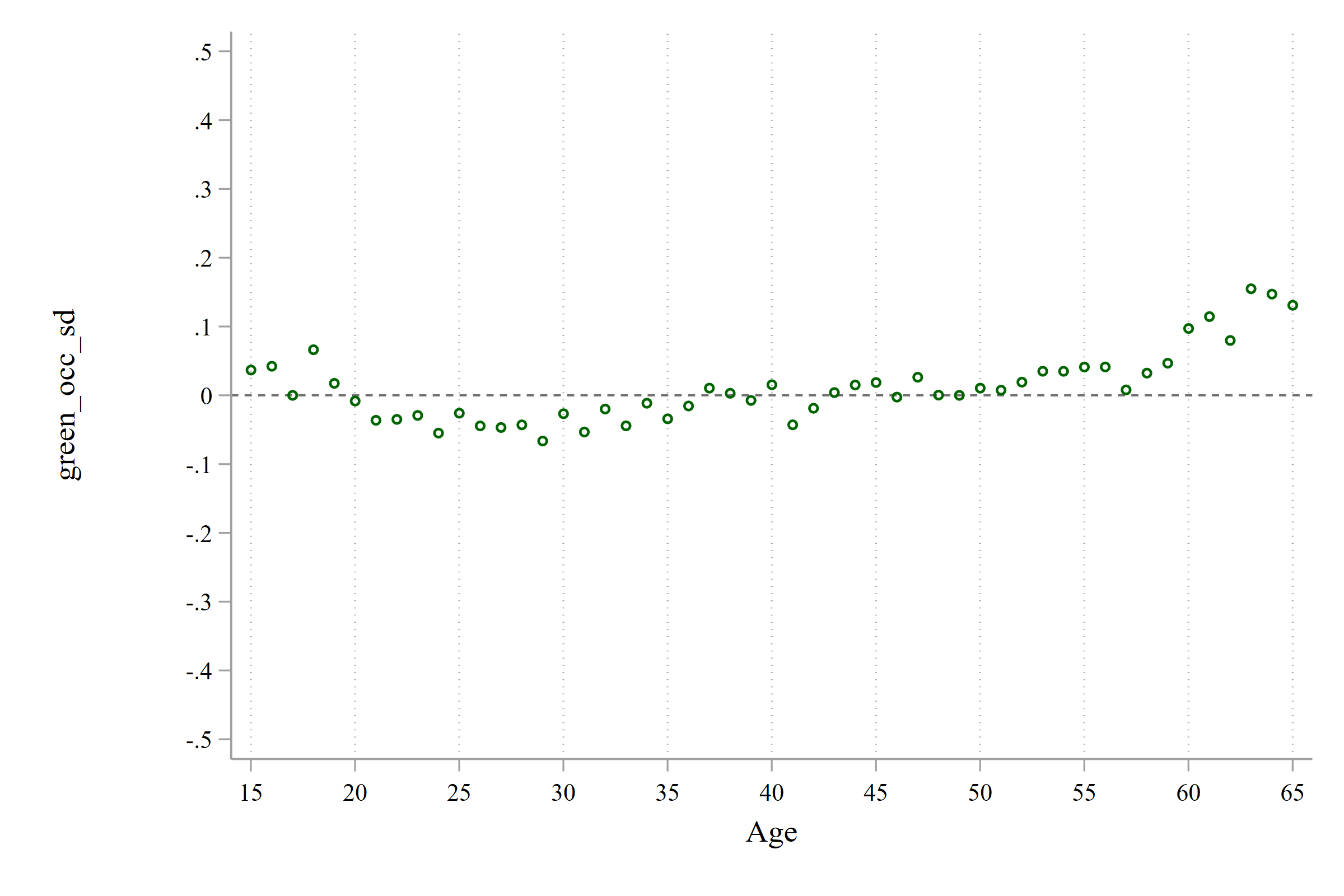}
     \fnote{{\smaller Own elaboration based on PHS. Figure shows the average green potential score across years of age. Survey weights were used. Calculations for the period 2015q1-2021q3. \textit{green\_occ} refers to a binary definition where an occupation is considered either green or non-green. The score was standardized such that for the entire sample of workers it has a mean of zero and a standard deviation of one.}}
         \label{fig:fig5}
\end{figure}

\begin{figure}[H]
    \centering
    \caption{Greenness over years of education}
    \includegraphics[width=\textwidth]{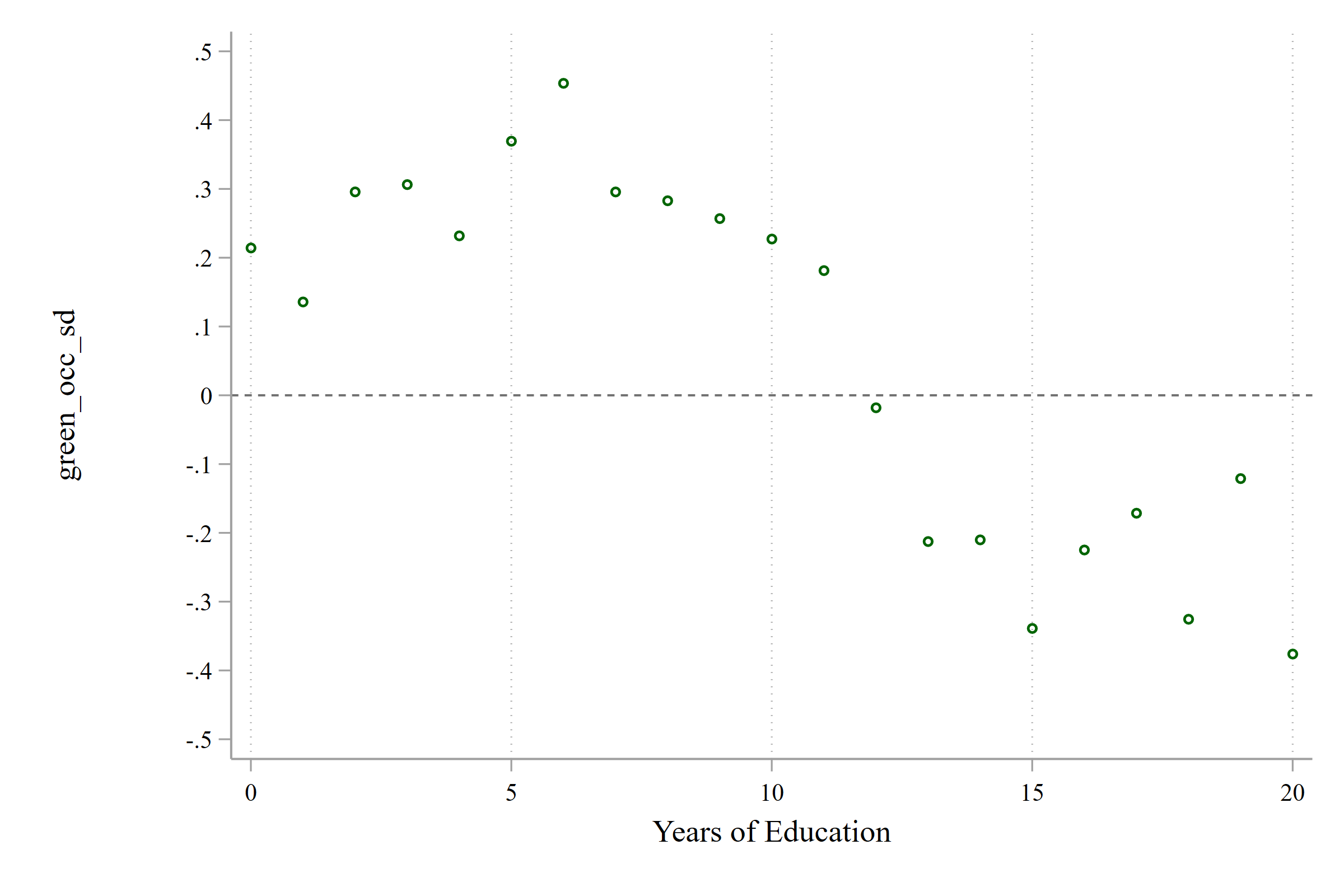}
     \fnote{{\smaller Own elaboration based on PHS. Figure shows the average green potential score across years of education. Survey weights were used. Calculations for the period 2015q1-2021q3. \textit{green\_occ} refers to a binary definition where an occupation is considered either green or non-green. The score was standardized such that for the entire sample of workers it has a mean of zero and a standard deviation of one.}}
         \label{fig:fig6}
\end{figure}

\begin{figure}[H]
    \centering
    \caption{Greenness over the income distribution}
    \includegraphics[width=\textwidth]{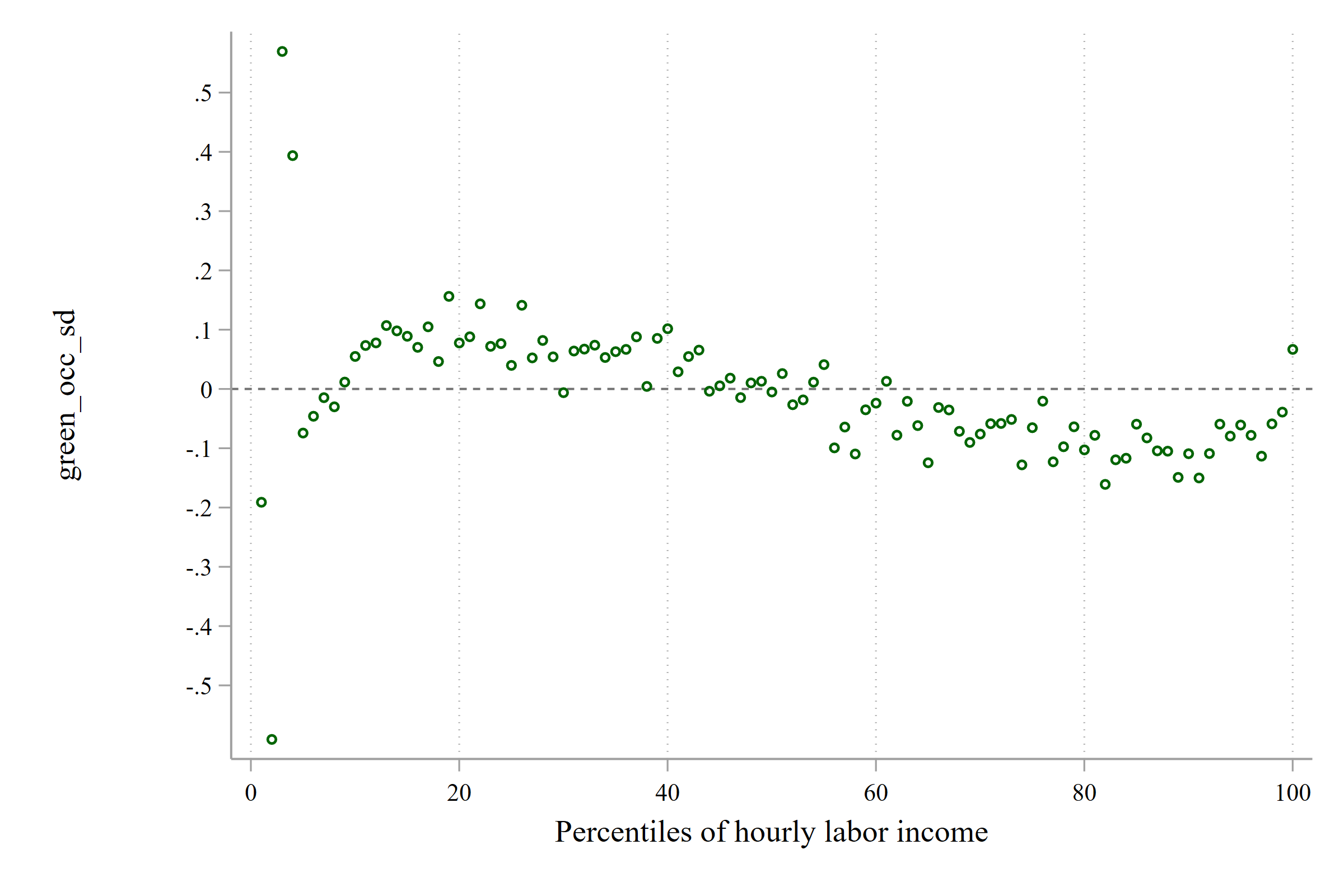}
     \fnote{{\smaller Own elaboration based on PHS. Figure shows the average green potential score across hourly wage percentiles. Survey weights were used. Calculations for the period 2015q1-2021q3. \textit{green\_occ} refers to a binary definition where an occupation is considered either green or non-green. The score was standardized such that for the entire sample of workers it has a mean of zero and a standard deviation of one.}}
         \label{fig:fig7}
\end{figure}

To explore how the green potential is distributed, Figure \ref{fig:fig5}, \ref{fig:fig6}, and \ref{fig:fig7} show the average greenness score, by years of age, years of education, and percentile of labor income, respectively. As for age, we found mean values close to zero until the age of 58, when it begins to slightly increase. In terms of the years of education, an inverted U-shaped relationship is evident, which initially increases until the age of 6, at which point it begins to decrease rapidly. As for labor income, the green potential is very close to zero across the entire income distribution.

\begin{figure}[H]
    \centering
    \caption{Greenness within sectors}
    \includegraphics[width=\textwidth]{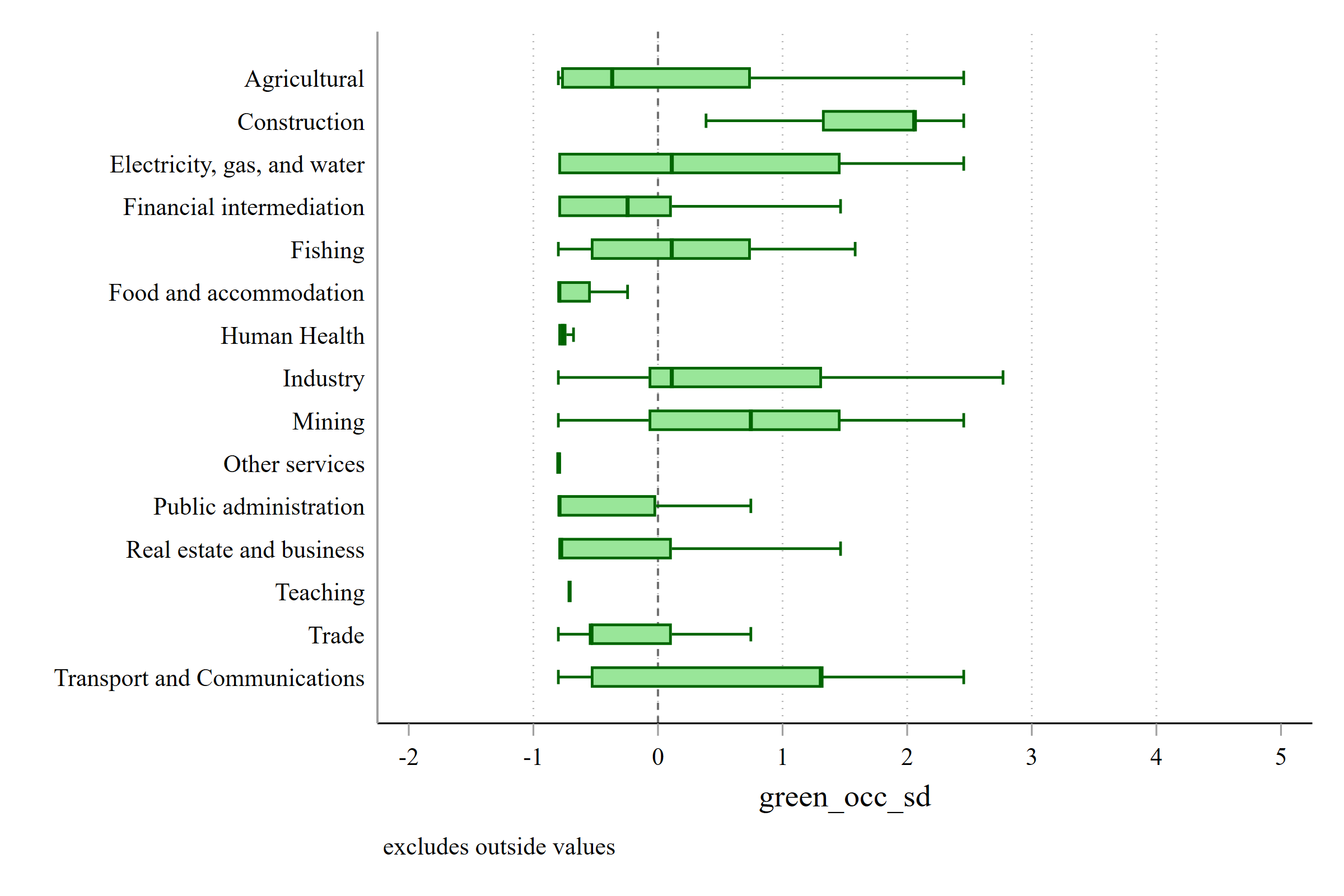}
     \fnote{{\smaller Own elaboration based on PHS. Survey weights were used. Calculations for the period 2015q1-2021q3. The black-green line is the median. \textit{green\_occ} refers to a binary definition where an occupation is considered either green or non-green. The score was standardized such that for the entire sample of workers it has a mean of zero and a standard deviation of one.}}
         \label{fig:fig8}
\end{figure}

In terms of economic sectors, construction, transportation and mining have the highest median green potential, while the median for industry, fishing and electricity, gas and water are very close to the sample median. The median green potential is very low in the rest of the sectors and reaches the minimum in human health, other services and education. However, there is significant heterogeneity within each sector (see Figure \ref{fig:fig8}).

\begin{figure}[H]
    \centering
    \caption{Greenness and informality (legal and productive)}
    Panel A: Legal Informality
  \includegraphics[width=\linewidth]{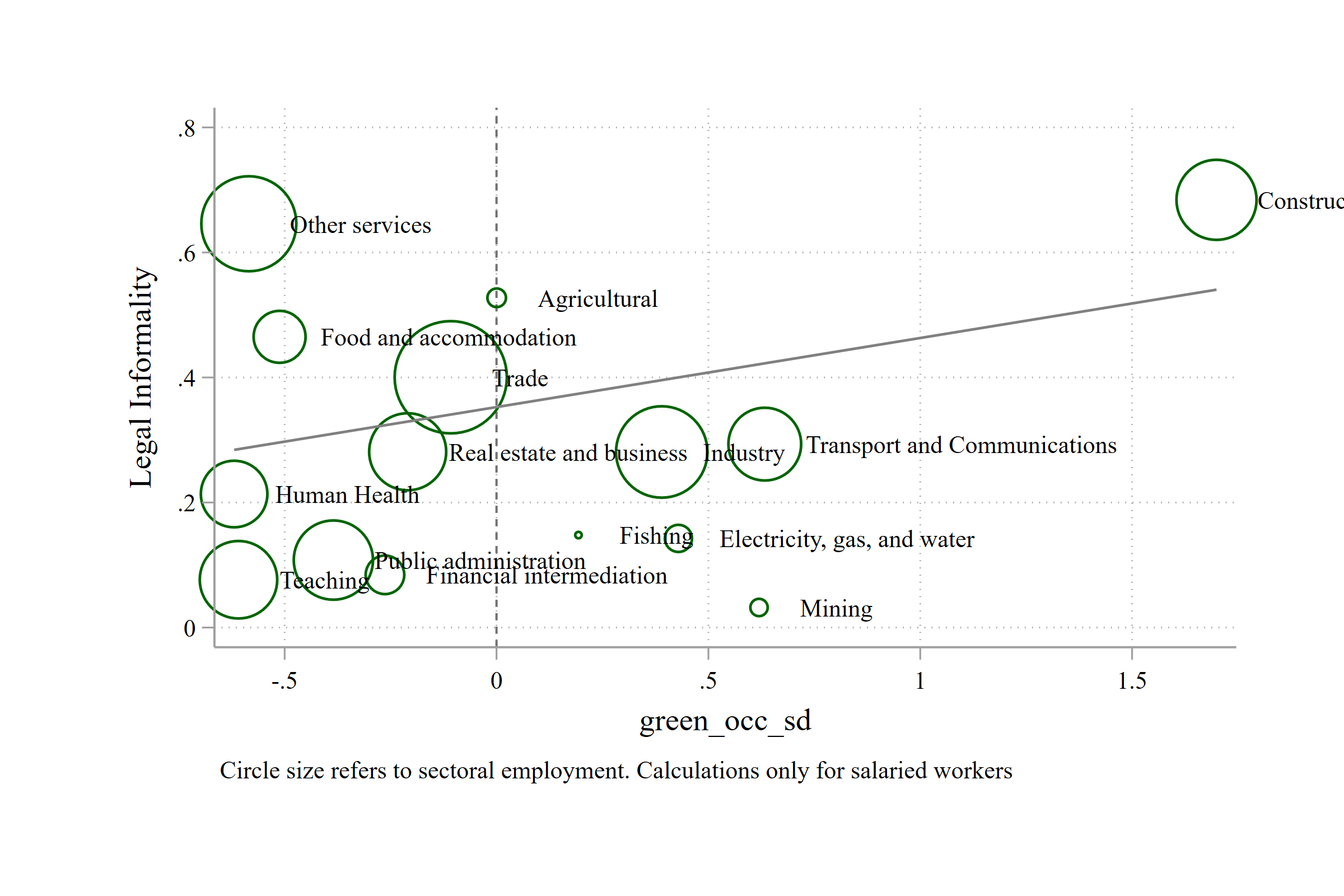}
    Panel B: Productive Informality
  \includegraphics[width=\linewidth]{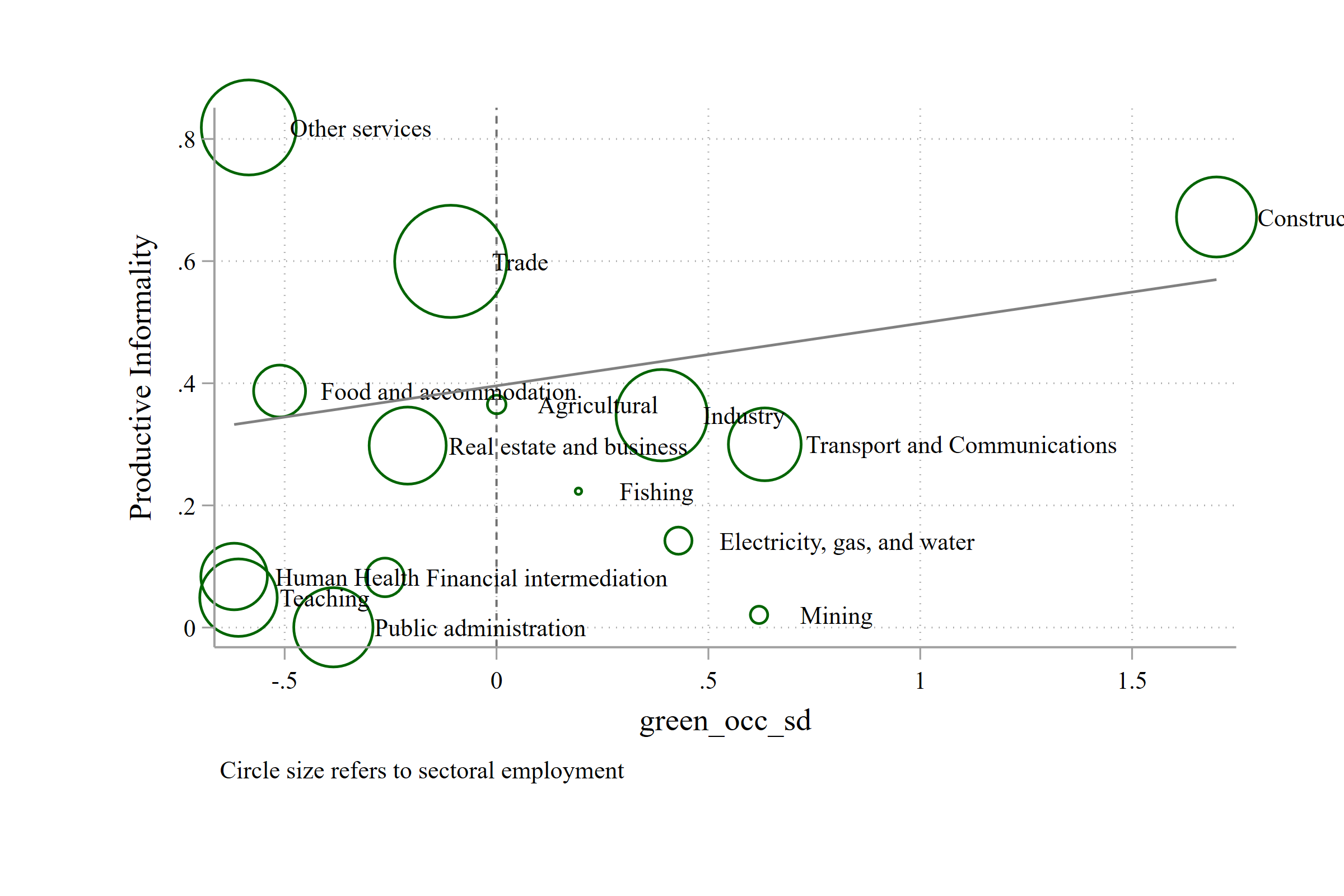}
     \fnote{{\smaller Own elaboration based on PHS. The figure shows average informality as a function of average green potential at the sectoral level. Survey weights were used. Calculations for the period 2015q1-2021q3. Legal informality refers to lacking social security rights, and can only be estimated for salaried workers. Productive informality identifies employees in small firms, non-professional self-employed workers, and workers without income. \textit{green\_occ} refers to a binary definition where an occupation is considered either green or non-green. The score was standardized such that for the entire sample of workers it has a mean of zero and a standard deviation of one.}}
         \label{fig:fig9}
\end{figure}

We are interested in how greenness indicators and various measures of decent work interact. As \cite{Ernst2019}, we use labor informality as a proxy for decent work. In addition, we use two different measures of informality, one that considers social security rights (legal informality) and another that bears into account productive conditions such as firm size, employment conditions, and workers without income (productive informality)\footnote{Productive informality identifies employees in small firms, non-professional self-employed workers, and workers without income.}\footnote{Note that legal informality can be calculated solely for wage earners.}. Figure \ref{fig:fig9} shows average informality as a function of average green potential at the sectoral level. There is no clear relationship between these two variables. Construction, transportation, and mining, the three sectors with the highest average green potential, evidence very different levels of labor informality. Something analogous occurs among the sectors with the lowest green potential: other services, human health, and teaching. For the rest of the sectors, with average green potential close to the sample mean, a negative relationship appears to exist, i.e., those with greater green potential have lower levels of informality.

\begin{figure}[H]
    \centering
    \caption{Greenness and Female Employment Share}
    \includegraphics[width=\textwidth]{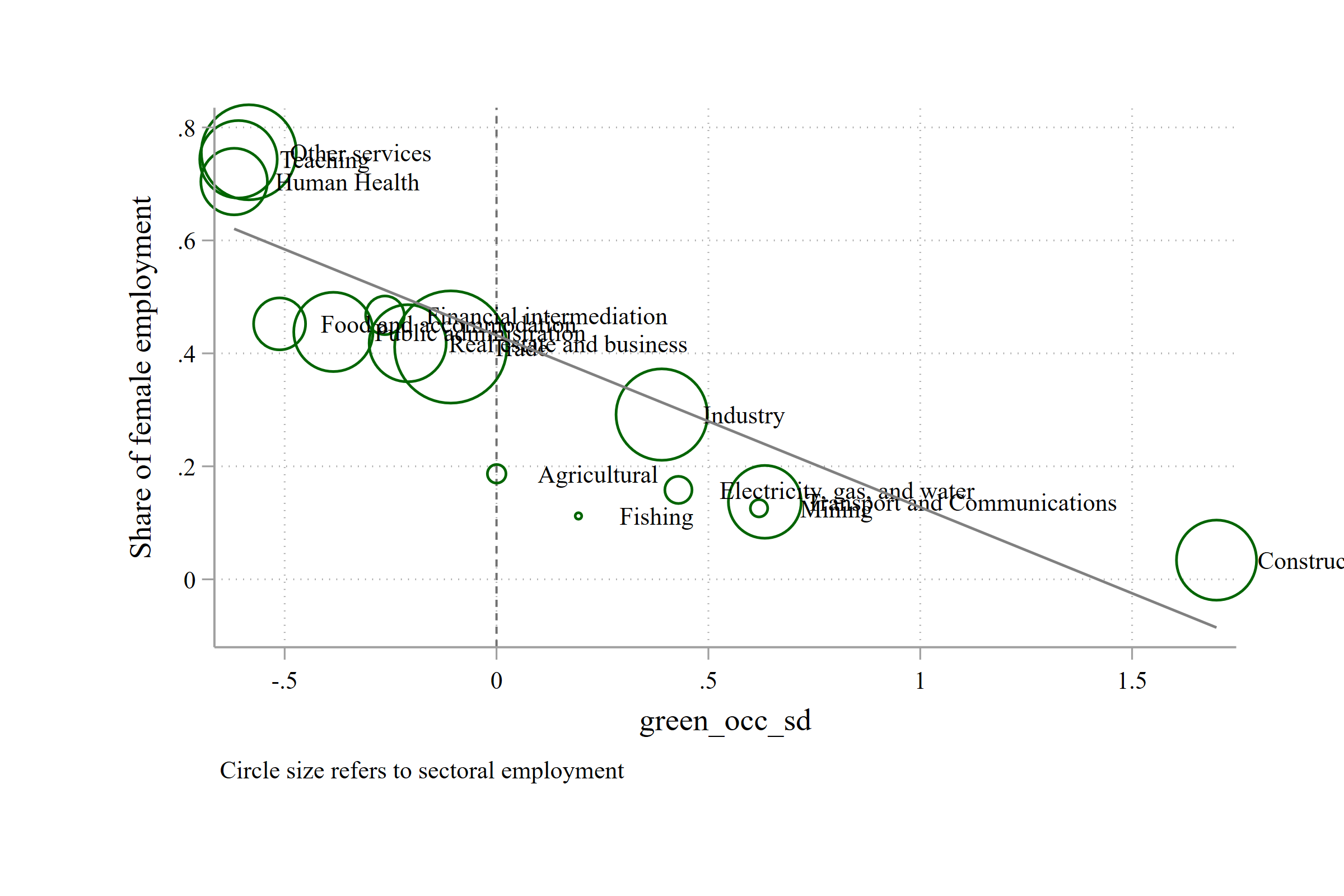}
     \fnote{{\smaller The figure shows female employment share as a function of average green potential at the sectoral level. Survey weights were used. Calculations for the period 2015q1-2021q3. \textit{green\_occ} refers to a binary definition where an occupation is considered either green or non-green. The score was standardized such that for the entire sample of workers it has a mean of zero and a standard deviation of one.}}
         \label{fig:fig10}
\end{figure}

A clearer pattern arises in terms of gender differences. As shown in Figure \ref{fig:fig10}, the green potential is lower in sectors with high feminization of employment such as teaching and human health, whereas it is higher in sectors with a low female employment share such as construction, transportation, and mining.

Table \ref{tab:tab1} shows the sectoral decomposition of green jobs, i.e., those with high green potential, and of formal green jobs (green jobs in formal positions)\footnote{See Section \ref{sec:lit_rev} for the definition of green jobs.}. We consider the two definitions of informality previously defined. We find that 25\% of workers are in green jobs. However, when taking into account the informality dimension, we find that between 15\% of workers and 12\% of wage earners are in formal green jobs (productive and legal informality, respectively).

\begin{table}[H]
  \caption{Green Jobs in the PHS, 2015q1-2021q3}

    \scalebox{0.72}{
     \begin{tabular}{|l|cc|ccc|}
    \multicolumn{1}{l}{\textit{Panel A: High Green Potential}} &       & \multicolumn{1}{r}{} &       &       & \multicolumn{1}{r}{} \\
    \midrule
    \multicolumn{1}{|c|}{\multirow{2}[4]{*}{Sector}} & \multicolumn{2}{c|}{Employment} & \multicolumn{3}{c|}{High green\_occ} \\
\cmidrule{2-6}          & thousands & \% of total & thousands & \% of sector & \% of total \\
    \midrule
    Agricultural & 48    & 0.47  & 11    & 21.74 & 0.41 \\
    Construction & 988   & 9.69  & 928   & 93.90 & 36.28 \\
    Electricity, gas, and water & 98    & 0.97  & 42    & 43.62 & 1.67 \\
    Financial intermediation & 209   & 2.05  & 5     & 2.62  & 0.22 \\
    Fishing & 5     & 0.05  & 1     & 22.85 & 0.05 \\
    Food and accommodation & 390   & 3.82  & 28    & 7.19  & 1.09 \\
    Human Health & 679   & 6.68  & 20    & 2.96  & 0.78 \\
    Industry & 1,16  & 11.41 & 366   & 31.53 & 14.32 \\
    Mining & 41    & 0.40  & 20    & 49.72 & 0.79 \\
    Other services & 1,383 & 13.56 & 92    & 6.70  & 3.61 \\
    Public administration & 869   & 8.55  & 87    & 10.04 & 3.44 \\
    Real estate and business & 760   & 7.47  & 109   & 14.39 & 4.29 \\
    Teaching & 871   & 8.56  & 15    & 1.71  & 0.59 \\
    Trade & 1,889 & 18.55 & 362   & 19.17 & 14.19 \\
    Transport and Communications & 789   & 7.75  & 467   & 59.23 & 18.28 \\
    \midrule
    Total & 10,18 & 100.00 & 2,553 & 25.07 & 100.00 \\
    \bottomrule
    \end{tabular}}
  \label{tab:tab1_panelA}%
  
  \bigskip
  \scalebox{0.72}{
     \begin{tabular}{|l|cc|ccc|ccc|}
    \multicolumn{9}{l}{\textit{Panel B: High Green Potential + Productive Formality}} \\
    \midrule
    \multicolumn{1}{|c|}{\multirow{2}[4]{*}{Sector}} & \multicolumn{2}{c|}{Employment} & \multicolumn{3}{c|}{Formal Employment} & \multicolumn{3}{c|}{High green\_occ + Formal} \\
\cmidrule{2-9}          & thousands & \% of total & thousands & \% of sector & \% of total & thousands & \% of sector & \% of total \\
    \midrule
    Agricultural & 46    & 0.47  & 14    & 44.59 & 0.29  & 11    & 23.14 & 0.71 \\
    Construction & 933   & 9.62  & 152   & 29.99 & 3.17  & 265   & 28.42 & 17.67 \\
    Electricity, gas, and water & 92    & 0.96  & 73    & 87.03 & 1.52  & 37    & 40.85 & 2.48 \\
    Financial intermediation & 196   & 2.03  & 174   & 93.12 & 3.63  & 16    & 7.94  & 1.05 \\
    Fishing & 5     & 0.05  & 3     & 85.92 & 0.07  & 1     & 23.94 & 0.08 \\
    Food and accommodation & 366   & 3.77  & 147   & 52.85 & 3.05  & 35    & 9.71  & 2.33 \\
    Human Health & 649   & 6.71  & 439   & 79.35 & 9.14  & 35    & 5.36  & 2.31 \\
    Industry & 1,078 & 11.15 & 544   & 70.71 & 11.32 & 272   & 25.26 & 18.14 \\
    Mining & 38    & 0.39  & 36    & 97.36 & 0.76  & 18    & 49.75 & 1.22 \\
    Other services & 1,351 & 13.93 & 372   & 35.04 & 7.76  & 45    & 3.32  & 2.98 \\
    Public administration & 868   & 8.98  & 773   & 89.12 & 16.10 & 100   & 11.48 & 6.68 \\
    Real estate and business & 692   & 7.14  & 348   & 72.31 & 7.26  & 118   & 17.10 & 7.89 \\
    Teaching & 846   & 8.75  & 741   & 93.11 & 15.44 & 47    & 5.60  & 3.19 \\
    Trade & 1,803 & 18.61 & 596   & 60.17 & 12.41 & 226   & 12.49 & 15.03 \\
    Transport and Communications & 721   & 7.44  & 389   & 70.72 & 8.09  & 273   & 37.96 & 18.24 \\
    \midrule
    Total & 9,683 & 100.00 & 4,801 & 66.61 & 100.00 & 1,499 & 15.48 & 100.00 \\
    \bottomrule
    \end{tabular}
     \label{tab:tab1_panelB}}
\label{tab:tab1}


  \bigskip
  \scalebox{0.72}{
  \begin{threeparttable}
\begin{tabular}{|l|cc|ccc|ccc|}
    \multicolumn{9}{l}{\textit{Panel C: High Green Potential + Legal Formality}} \\
    \midrule
    \multicolumn{1}{|c|}{\multirow{2}[4]{*}{Sector}} & \multicolumn{2}{c|}{Employment} & \multicolumn{3}{c|}{Formal Employment} & \multicolumn{3}{c|}{High green\_occ + Formal} \\
\cmidrule{2-9}          & thousands & \% of total & thousands & \% of sector & \% of total & thousands & \% of sector & \% of total \\
    \midrule
    Agricultural & 33    & 0.43  & 15    & 44.63 & 0.29  & 4     & 12.84 & 0.44 \\
    Construction & 566   & 7.32  & 173   & 30.80 & 3.39  & 142   & 25.23 & 14.74 \\
    Electricity, gas, and water & 90    & 1.17  & 78    & 86.97 & 1.53  & 37    & 40.78 & 3.80 \\
    Financial intermediation & 200   & 2.60  & 185   & 92.38 & 3.62  & 12    & 6.14  & 1.29 \\
    Fishing & 4     & 0.06  & 4     & 85.34 & 0.07  & 1     & 25.29 & 0.12 \\
    Food and accommodation & 303   & 3.91  & 158   & 52.61 & 3.10  & 10    & 3.42  & 1.03 \\
    Human Health & 583   & 7.59  & 459   & 78.79 & 9.00  & 17    & 2.85  & 1.73 \\
    Industry & 852   & 11.03 & 596   & 69.92 & 11.65 & 216   & 25.45 & 22.42 \\
    Mining & 40    & 0.53  & 39    & 96.81 & 0.77  & 20    & 50.90 & 2.11 \\
    Other services & 1,107 & 14.31 & 390   & 35.65 & 7.65  & 23    & 2.12  & 2.44 \\
    Public administration & 869   & 11.28 & 774   & 89.10 & 15.16 & 87    & 9.97  & 9.04 \\
    Real estate and business & 551   & 7.15  & 398   & 72.39 & 7.81  & 35    & 6.35  & 3.64 \\
    Teaching & 820   & 10.65 & 762   & 92.87 & 14.93 & 43    & 5.19  & 4.45 \\
    Trade & 1,077 & 13.95 & 643   & 59.72 & 12.59 & 102   & 9.50  & 10.63 \\
    Transport and Communications & 619   & 8.02  & 431   & 69.72 & 8.44  & 213   & 34.48 & 22.14 \\
    \midrule
    Total & 7,716 & 100.00 & 5,105 & 66.26 & 100.00 & 961   & 12.48 & 100.00 \\
    \bottomrule
\end{tabular}
     \begin{tablenotes}
    \small \item Own elaboration based on PHS. Panel A shows the share of workers with high green potential in the PHS. Panel B shows the share of workers with high green potential who are in formal positions (productive definition) in the PHS. \textit{green\_occ} refers to a binary definition where an occupation is considered either green or non-green. The scores were standardized such that for the entire sample of workers, each greenness measure has a mean of zero and a standard deviation of one. Survey weights were used.
\end{tablenotes}
  \end{threeparttable}
  \label{tab:tab1_panelC}}
\end{table}%

\section{Empirical Strategy and Results} \label{sec:results}

To explore how the greenness score is distributed, we estimate the relationship between the greenness score and various labor and demographic variables at the individual level by estimating the following equation:
\begin{align}
greennes_{i}=b_{0}+b_{1} X_{i}+reg+quarter+e_{i}
\end{align}
where $greennes_{i}$ is either a standardized greenness score or an indicator variable of green jobs/formal green jobs based on the individual's $i$ occupation; $X_i$ is a set of demographic and employment characteristics. The vector $X_i$ includes gender, age, educational level, the activity sector, the type and size of the company, the income decile of the main occupation, and a dummy variable indicating whether the position is informal. Finally, we include time fixed effects by quarter, quarter, and regional fixed effects by urban area, $reg$; and $e_i$ is the unobservable error term. The period under analysis is from 2015q1 to 2021q3. Descriptive statistics of both the dependent variables and the independent variables are reported in Table \ref{tab:table2}, and \ref{tab:table3}, respectively. 

We cluster the standard errors at the ISCO-08 2-digit occupation level, that is, at the same level as the green potential variables, to consider the possible correlation between unobservable characteristics of individuals employed in the same occupation. Alternatively, the estimates were repeated using robust standard errors and the significance levels are practically identical\footnote{Results are available upon request.}. For each dependent variable we ran three specifications: one for the full sample, one for the subsample of wage earners and using the productive definition of informality as independent variable, and the third for the subsample of wage earners and using the legal definition of informality as independent variable\footnote{Since \textit{high\_green\_occ\_formal} can be calculated solely for wage earners, there are two specifications for this case.}. Table 4 (in the Appendix) presents the results.

\begin{table}[H]
  \centering
  \caption{Summary statistics. Greenness indicators}
  \begin{threeparttable}
     \begin{tabular}{lcccccccc}
    \toprule
    \multicolumn{1}{l}{Green Indicador} & \multicolumn{1}{c}{N} & \multicolumn{1}{c}{mean} & \multicolumn{1}{c}{sd} & \multicolumn{1}{c}{min} & \multicolumn{1}{c}{p25} & \multicolumn{1}{c}{p50} & \multicolumn{1}{c}{p75} & \multicolumn{1}{c}{max} \\
    \midrule
    green\_occ\_sd & 383,865 & 0.00  & 1.02  & -0.80 & -0.80 & -0.54 & 1.31  & 2.77 \\
    high\_green\_occ & 383,865 & 0.26  & 0.44  & 0.00  & 0.00  & 0.00  & 1.00  & 1.00 \\
    high\_green\_occ\_prform & 383,865 & 0.13  & 0.34  & 0.00  & 0.00  & 0.00  & 0.00  & 1.00 \\
    high\_green\_occ\_formal & 289,424 & 0.11  & 0.31  & 0.00  & 0.00  & 0.00  & 0.00  & 1.00 \\
    \bottomrule
    \end{tabular}
     \begin{tablenotes}
    \smaller \item Own elaboration. Survey weights were used. \textit{green\_occ} refers to a binary definition where an occupation is considered either green or non-green; \textit{high\_green\_occ} is a dummy variable that equals 1 for those occupations with a green potential above 2.5 times the mean, and 0 otherwise; \textit{high\_green\_occ\_prform} and \textit{high\_green\_occ\_formal} are the same as before but also equals 0 for those in informal position (productive and legal definition, respectively).
\end{tablenotes}
  \end{threeparttable}
  \label{tab:table2}%
\end{table}%

\begin{ThreePartTable}
\footnotesize\setlength{\tabcolsep}{2.6pt}
\begin{TableNotes}
    \smaller \item Own elaboration. Survey weights were used. \textit{green\_occ} refers to a binary definition where an occupation is considered either green or non-green; \textit{greenness} is a task-based indicator that accounts for the proportion of green tasks on total tasks within an occupation; and \textit{tgreenness\_core} is a task-based indicator that accounts for the proportion of green core tasks on total core tasks within an occupation.
\end{TableNotes}
\begin{longtable}{lccc}
  \caption{Summary statistics. Independent Variables}   \label{tab:table3} \\
  
    \toprule
          & \multicolumn{1}{c}{N} & \multicolumn{1}{c}{mean} & \multicolumn{1}{c}{sd} \\ \hline 
\endfirsthead

\multicolumn{3}{c}%
{{\bfseries \tablename\ \thetable{} -- continued from previous page}} \\
\toprule & \multicolumn{1}{c}{N} & \multicolumn{1}{c}{mean} & \multicolumn{1}{c}{sd} \\ \hline  
\endhead
\endfoot

\bottomrule \addlinespace
\insertTableNotes
\endlastfoot

    \midrule
    \textbf{Men} & 383,865 &                0.56  &                0.50  \\
    \multicolumn{4}{l}{\textbf{Age groups}} \\
    $[15,24]$ & 383,865 &                0.12  &                0.32  \\
    $[25,34]$ & 383,865 &                0.26  &                0.44  \\
    $[35,44]$ & 383,865 &                0.27  &                0.44  \\
    $[45,54]$ & 383,865 &                0.21  &                0.41  \\
    $[55,64]$ & 383,865 &                0.13  &                0.34  \\
    $[65+]$ & 383,865 &                0.01  &                0.08  \\
    \multicolumn{4}{l}{\textbf{Region}} \\
    Gran Buenos Aires & 383,865 &                0.14  &                0.35  \\
    Noroeste & 383,865 &                0.25  &                0.43  \\
    Noreste & 383,865 &                0.10  &                0.30  \\
    Cuyo  & 383,865 &                0.11  &                0.32  \\
    Pampeana & 383,865 &                0.28  &                0.45  \\
    Patagonia & 383,865 &                0.13  &                0.33  \\
    \multicolumn{4}{l}{\textbf{Educational level}} \\
    Never attended & 383,865 &                0.00  &                0.06  \\
    Primary incomplete & 383,865 &                0.03  &                0.18  \\
    Primary complete & 383,865 &                0.15  &                0.35  \\
    Secondary incomplete & 383,865 &                0.18  &                0.38  \\
    Secondary complete & 383,865 &                0.30  &                0.46  \\
    Superior incomplete & 383,865 &                0.14  &                0.34  \\
    Superior complete & 383,865 &                0.20  &                0.40  \\
    \multicolumn{4}{l}{\textbf{Sector}} \\
    Agricultural & 383,865 &                0.01  &                0.09  \\
    Construction & 383,865 &                0.11  &                0.32  \\
    Electricity, gas, and water & 383,865 &                0.01  &                0.10  \\
    Financial intermediation & 383,865 &                0.02  &                0.12  \\
    Fishing & 383,865 &                0.00  &                0.02  \\
    Food and accommodation & 383,865 &                0.04  &                0.19  \\
    Human Health & 383,865 &                0.06  &                0.24  \\
    Industry & 383,865 &                0.10  &                0.29  \\
    Mining & 383,865 &                0.01  &                0.08  \\
    Other services & 383,865 &                0.14  &                0.35  \\
    Public administration & 383,865 &                0.12  &                0.32  \\
    Real estate and business & 383,865 &                0.06  &                0.24  \\
    Teaching & 383,865 &                0.08  &                0.28  \\
    Trade & 383,865 &                0.19  &                0.39  \\
    Transport and Communications & 383,865 &                0.06  &                0.24  \\
    \multicolumn{4}{l}{\textbf{Firm Type}} \\
    Public & 383,865 &                0.22  &                0.42  \\
    Private & 383,865 &                0.77  &                0.42  \\
    Other & 383,865 &                0.01  &                0.10  \\
    \textbf{Informal (productive definition)} & 383,865 &                0.42  &                0.49  \\
    \textbf{Informal (legal definition)} & 289,424 &                0.36  &                0.48  \\
    \multicolumn{4}{l}{\textbf{Firm Size}} \\
    5 or less employees & 383,865 &                0.48  &                0.50  \\
    between 6 and 40 employees & 383,865 &                0.26  &                0.44  \\
    more than 40 & 383,865 &                0.26  &                0.44  \\
    \multicolumn{4}{l}{\textbf{Hourly labor income decile}} \\
    1st decile & 383,865 &                0.10  &                0.31  \\
    2nd decile & 383,865 &                0.12  &                0.32  \\
    3rd decile & 383,865 &                0.11  &                0.31  \\
    4th decile & 383,865 &                0.10  &                0.31  \\
    5th decile & 383,865 &                0.10  &                0.30  \\
    6th decile & 383,865 &                0.10  &                0.30  \\
    7th decile & 383,865 &                0.10  &                0.29  \\
    8th decile & 383,865 &                0.09  &                0.29  \\
    9th decile & 383,865 &                0.09  &                0.29  \\
    10th decile & 383,865 &                0.08  &                0.28  \\
    \bottomrule
\end{longtable}
\end{ThreePartTable}

First, we find that men have higher green potential than women, but this gap narrows when using green formal jobs as the dependent variable (see columns 7-11). In relation to age, we find a clear increasing relationship but solely for the full sample under the \textit{green\_occ} score definition (column 1). In regional terms, those who reside in the Pampeana and Patagonian regions show a green potential greater than GBA, while there are no statistically significant differences in relation to the rest of the regions. There does not appear to be a statistically significant relationship between educational level and greening variables, except for those with completed higher education, who have a higher greening potential. Taking agricultural activities as the base category, the sectors with the highest green potential are construction, transportation, mining, and industry, whereas human health has significantly lower green potential. The green potential is decreasing in the firm size, except for the case of green formal jobs (legal definition) for which the opposite is true. There is no statistically significant relationship between the green potential and the income decile of the main occupation. Being an informal worker, according to the productive definition, is negatively and significantly related to the four measures of green potential (see columns 1, 4, 7, and 10). However, we find that being in informal jobs is positively associated with the probability of being in a high green potential occupation for the subsample of wage earners (see column 5). Similarly, when using the legal definition of informality, we find that it is positively related with both the green potential and with the probability of being in a high green potential occupation (see columns 3 and 6). By construction, it is reasonable to find a negative relationship between both informality variables and the probability of being in green formal jobs (columns 7-11). 

In summary, the groups that are likely to benefit most by the greening of the Argentine economy are those characterized by being men, elderly, those with very high qualifications (higher education), and those in specific sectors such as construction, transportation, mining, and industry. These results are aligned with previous findings for Argentina \cite{Ernst2019} and for other countries including the US and several European economies \citep{Consoli2016, Vona2019, Lobsiger2021, Valero2021}. At the same time, the specific sectors identified as benefited are in line with some of the major green economy sectors reviewed by \cite{Dierdorff2009}. The most striking result is that while informality is negatively associated with the green potential in the full sample of workers, the opposite is true for the subsample of wage earners, especially if we consider the broader definition of green potential (\textit{green\_occ\_sd}) and a more restrictive one (\textit{high\_green\_occ}). Therefore, the green transition may be incompatible with decent job search for the latter.

\section{Conclusion} \label{sec:conclusion}

Several countries, including Argentina, recognize the importance of greening the economy and have been developing strategies to foster the implementation of green technologies, the creation of sustainable industries, the reduction in current pollution, among other goals. Despite a broad consensus over the expected long-term benefits of such transition, in the short/medium run some frictions and those workers who perform tasks and use skills that are incompatible with the new labor market conditions will be the most affected. In addition, the ILO’s Green Jobs initiative recognizes the importance of both the creation of decent work in the new green activities and the implementation of social protection policies to mitigate the effects on the sectors that need transforming to ensure a more inclusive green transition.

Defining the green potential of jobs as the ability of an occupation to perform green tasks and taking into account informality as a proxy of decent work, this paper characterizes the potential of green jobs in Argentina for the period 2015-2021. Our paper is the first to apply an occupation-based approach with a greenness perspective in a developing country like Argentina, which faces not only difficulties related to the changes in labor skills and production methods that a green transition would require, but also to the fact that it has a considerable proportion of informal employment, a common feature among developing economies.

Our results reveal that 25\% of workers are currently in green jobs and 15\% are in formal green jobs (approximately 60\% of workers that are in green jobs). At the same time, the green potential is relatively higher for men, the elderly, those with completed higher education, and those in specific sectors such as construction, transportation, mining, and industry. All in all, these results suggest that it is easier to transition to greener jobs for workers in those groups. Furthermore, in a country like Argentina, the informality dimension is crucial. Our results suggest that for the full sample of workers the informality is negatively associated with all the green potential measures. Nevertheless, if we restrict the sample to wage earners, we find that the opposite holds true: informality is positively associated with both the green potential and the high green potential variable. Therefore, those workers in formal jobs are also likely to benefit from the green transition, although for wage earners this transition may be incompatible with decent work.   

Policies will be needed to efficiently manage the transition, i.e., to maximize the gains of those who benefit and support those who would be harmed in the short/medium run.  For instance, the \cite{IMF2022} proposes some policies such as job training, tax credits for lower-income workers, promotion of investment in R\&D and green infrastructure, and a carbon tax. Many countries, especially the developing and emerging ones like Argentina, could take advantage of this likely change in its economic structure and could seek to implement some of the recommended labor policies with the hope that it will also help boost the informal workers out of that condition.   
The limitations of our study are crystal clear since our measurements of occupational greenness are calculated with data from the United States, given that Argentina does not have descriptions of work content and skills at the occupational level. Further investigation will be necessary when data availability ceases to be a constraint. In addition, further research could be focused on studying the effects of several environmental policies that Argentina is currently undertaking. It would be somehow useful to address the consequences of such policies on several variables, such as the effect on the labor market or on the economy as a whole, and investigate whether they are effective in achieving the established goals. One line of research could be to examine if these policies are associated with a heterogeneous change in the demand for workers with higher green potential relative to those with lower green potential.


\clearpage

\bibliography{bibliography}

\pagebreak


\appendix
\begin{landscape}
\section*{Appendix} \label{sec:appendix}

\tiny
\setlength\LTcapwidth{\textwidth} 
\setlength\LTleft{0pt}
\setlength\LTright{0pt}

\begin{ThreePartTable}
\footnotesize\setlength{\tabcolsep}{2.6pt}
\begin{TableNotes}
    \smaller \item Estimates obtained by OLS. Sample weights are used. Cluster-corrected standard errors at the 2-digit level of ISCO-08 in parentheses. Quarter fixed effects included. *** p $<$0.01, ** p $<$0.05, * p $<$0.1. \textit{green\_occ} refers to a binary definition where an occupation is considered either green or non-green; \textit{high\_green\_occ} is a dummy variable that equals 1 for those occupations with a green potential above 2.5 times the mean, and 0 otherwise; \textit{high\_green\_occ\_prform} and \textit{high\_green\_occ\_formal} are the same as before but also equals 0 for those in informal position (productive and legal definition, respectively).
\end{TableNotes}
\setlength{\tabcolsep}{1pt}

\begin{longtable}{lrrrrrrrrrrr}
\caption{Estimation results}
\label{tab:table4} \\

\toprule
\multicolumn{1}{c}{} & \multicolumn{1}{c}{(1)} & \multicolumn{1}{c}{(2)} & \multicolumn{1}{c}{(3)} & \multicolumn{1}{c}{(4)} & \multicolumn{1}{c}{(5)} & \multicolumn{1}{c}{(6)} & \multicolumn{1}{c}{(7)} & \multicolumn{1}{c}{(8)} & \multicolumn{1}{c}{(9)} & \multicolumn{1}{c}{(10)} & \multicolumn{1}{c}{(11)} \\
          & \multicolumn{3}{c}{green\_occ\_sd} & \multicolumn{3}{c}{high\_green\_occ} & \multicolumn{3}{c}{high\_green\_occ\_prform} & \multicolumn{2}{c}{high\_green\_occ\_formal} \\
 & \multicolumn{1}{c}{Full sample} & \multicolumn{1}{c}{Wage earners} & \multicolumn{1}{c}{Wage earners} & \multicolumn{1}{c}{Full sample} & \multicolumn{1}{c}{Wage earners} & \multicolumn{1}{c}{Wage earners} & \multicolumn{1}{c}{Full sample} & \multicolumn{1}{c}{Wage earners} & \multicolumn{1}{c}{Wage earners} & \multicolumn{1}{c}{Wage earners} & \multicolumn{1}{c}{Wage earners} \\

    \hline 
\endfirsthead

 \\
\toprule
    \multicolumn{1}{c}{} & \multicolumn{1}{c}{(1)} & \multicolumn{1}{c}{(2)} & \multicolumn{1}{c}{(3)} & \multicolumn{1}{c}{(4)} & \multicolumn{1}{c}{(5)} & \multicolumn{1}{c}{(6)} & \multicolumn{1}{c}{(7)} & \multicolumn{1}{c}{(8)} & \multicolumn{1}{c}{(9)} & \multicolumn{1}{c}{(10)} & \multicolumn{1}{c}{(11)} \\
 \hline  
\endhead
\endfoot

\bottomrule \addlinespace
\insertTableNotes
\endlastfoot

\midrule
   VARIABLES &       &       &       &       &       &       &       &       &       &       &  \\
    =1 if male = 1 & \multicolumn{1}{c}{0.377***} & \multicolumn{1}{c}{0.349***} & \multicolumn{1}{c}{0.351***} & \multicolumn{1}{c}{0.173***} & \multicolumn{1}{c}{0.158***} & \multicolumn{1}{c}{0.159***} & \multicolumn{1}{c}{0.066***} & \multicolumn{1}{c}{0.086**} & \multicolumn{1}{c}{0.087**} & \multicolumn{1}{c}{0.087***} & \multicolumn{1}{c}{0.082**} \\
          & \multicolumn{1}{c}{(0.119)} & \multicolumn{1}{c}{(0.110)} & \multicolumn{1}{c}{(0.110)} & \multicolumn{1}{c}{(0.058)} & \multicolumn{1}{c}{(0.053)} & \multicolumn{1}{c}{(0.053)} & \multicolumn{1}{c}{(0.024)} & \multicolumn{1}{c}{(0.032)} & \multicolumn{1}{c}{(0.032)} & \multicolumn{1}{c}{(0.032)} & \multicolumn{1}{c}{(0.030)} \\
    \multicolumn{5}{l}{Age groups (base = $[15,24]$)} &       &       &       &       &       &       &  \\
    $[25,34]$ & \multicolumn{1}{c}{0.043} & \multicolumn{1}{c}{0.020} & \multicolumn{1}{c}{0.028} & \multicolumn{1}{c}{0.005} & \multicolumn{1}{c}{0.000} & \multicolumn{1}{c}{0.004} & \multicolumn{1}{c}{0.010} & \multicolumn{1}{c}{0.013} & \multicolumn{1}{c}{0.014} & \multicolumn{1}{c}{0.031**} & \multicolumn{1}{c}{0.012} \\
          & \multicolumn{1}{c}{(0.037)} & \multicolumn{1}{c}{(0.038)} & \multicolumn{1}{c}{(0.037)} & \multicolumn{1}{c}{(0.025)} & \multicolumn{1}{c}{(0.023)} & \multicolumn{1}{c}{(0.022)} & \multicolumn{1}{c}{(0.012)} & \multicolumn{1}{c}{(0.013)} & \multicolumn{1}{c}{(0.014)} & \multicolumn{1}{c}{(0.014)} & \multicolumn{1}{c}{(0.010)} \\
    $[35,44]$ & \multicolumn{1}{c}{0.085*} & \multicolumn{1}{c}{0.042} & \multicolumn{1}{c}{0.055} & \multicolumn{1}{c}{0.014} & \multicolumn{1}{c}{0.006} & \multicolumn{1}{c}{0.012} & \multicolumn{1}{c}{0.019} & \multicolumn{1}{c}{0.023} & \multicolumn{1}{c}{0.025} & \multicolumn{1}{c}{0.047**} & \multicolumn{1}{c}{0.018} \\
          & \multicolumn{1}{c}{(0.047)} & \multicolumn{1}{c}{(0.049)} & \multicolumn{1}{c}{(0.048)} & \multicolumn{1}{c}{(0.032)} & \multicolumn{1}{c}{(0.031)} & \multicolumn{1}{c}{(0.030)} & \multicolumn{1}{c}{(0.016)} & \multicolumn{1}{c}{(0.019)} & \multicolumn{1}{c}{(0.019)} & \multicolumn{1}{c}{(0.019)} & \multicolumn{1}{c}{(0.012)} \\
    $[45,54]$ & \multicolumn{1}{c}{0.101*} & \multicolumn{1}{c}{0.054} & \multicolumn{1}{c}{0.069} & \multicolumn{1}{c}{0.014} & \multicolumn{1}{c}{0.006} & \multicolumn{1}{c}{0.013} & \multicolumn{1}{c}{0.021} & \multicolumn{1}{c}{0.029} & \multicolumn{1}{c}{0.032} & \multicolumn{1}{c}{0.051**} & \multicolumn{1}{c}{0.016} \\
          & \multicolumn{1}{c}{(0.053)} & \multicolumn{1}{c}{(0.057)} & \multicolumn{1}{c}{(0.056)} & \multicolumn{1}{c}{(0.036)} & \multicolumn{1}{c}{(0.036)} & \multicolumn{1}{c}{(0.035)} & \multicolumn{1}{c}{(0.019)} & \multicolumn{1}{c}{(0.023)} & \multicolumn{1}{c}{(0.023)} & \multicolumn{1}{c}{(0.020)} & \multicolumn{1}{c}{(0.012)} \\
    $[55,64]$ & \multicolumn{1}{c}{0.112**} & \multicolumn{1}{c}{0.048} & \multicolumn{1}{c}{0.063} & \multicolumn{1}{c}{0.008} & \multicolumn{1}{c}{-0.004} & \multicolumn{1}{c}{0.004} & \multicolumn{1}{c}{0.014} & \multicolumn{1}{c}{0.024} & \multicolumn{1}{c}{0.026} & \multicolumn{1}{c}{0.044**} & \multicolumn{1}{c}{0.009} \\
          & \multicolumn{1}{c}{(0.054)} & \multicolumn{1}{c}{(0.058)} & \multicolumn{1}{c}{(0.055)} & \multicolumn{1}{c}{(0.037)} & \multicolumn{1}{c}{(0.035)} & \multicolumn{1}{c}{(0.033)} & \multicolumn{1}{c}{(0.017)} & \multicolumn{1}{c}{(0.020)} & \multicolumn{1}{c}{(0.020)} & \multicolumn{1}{c}{(0.020)} & \multicolumn{1}{c}{(0.014)} \\
    $[65+]$ & \multicolumn{1}{c}{0.136**} & \multicolumn{1}{c}{0.049} & \multicolumn{1}{c}{0.057} & \multicolumn{1}{c}{0.013} & \multicolumn{1}{c}{-0.010} & \multicolumn{1}{c}{-0.006} & \multicolumn{1}{c}{0.011} & \multicolumn{1}{c}{0.025} & \multicolumn{1}{c}{0.026} & \multicolumn{1}{c}{0.032*} & \multicolumn{1}{c}{0.013} \\
          & \multicolumn{1}{c}{(0.064)} & \multicolumn{1}{c}{(0.072)} & \multicolumn{1}{c}{(0.070)} & \multicolumn{1}{c}{(0.040)} & \multicolumn{1}{c}{(0.040)} & \multicolumn{1}{c}{(0.039)} & \multicolumn{1}{c}{(0.018)} & \multicolumn{1}{c}{(0.023)} & \multicolumn{1}{c}{(0.022)} & \multicolumn{1}{c}{(0.018)} & \multicolumn{1}{c}{(0.017)} \\
    \multicolumn{5}{l}{Region (base = Gran Buenos Aires)} &       &       &       &       &       &       &  \\
    Noroeste & \multicolumn{1}{c}{0.002} & \multicolumn{1}{c}{-0.011} & \multicolumn{1}{c}{-0.013} & \multicolumn{1}{c}{0.008} & \multicolumn{1}{c}{0.004} & \multicolumn{1}{c}{0.004} & \multicolumn{1}{c}{0.007} & \multicolumn{1}{c}{-0.001} & \multicolumn{1}{c}{-0.001} & \multicolumn{1}{c}{0.004} & \multicolumn{1}{c}{0.008} \\
          & \multicolumn{1}{c}{(0.013)} & \multicolumn{1}{c}{(0.014)} & \multicolumn{1}{c}{(0.014)} & \multicolumn{1}{c}{(0.005)} & \multicolumn{1}{c}{(0.006)} & \multicolumn{1}{c}{(0.006)} & \multicolumn{1}{c}{(0.008)} & \multicolumn{1}{c}{(0.007)} & \multicolumn{1}{c}{(0.007)} & \multicolumn{1}{c}{(0.007)} & \multicolumn{1}{c}{(0.007)} \\
    Noreste & \multicolumn{1}{c}{-0.009} & \multicolumn{1}{c}{-0.037*} & \multicolumn{1}{c}{-0.037*} & \multicolumn{1}{c}{0.003} & \multicolumn{1}{c}{-0.006} & \multicolumn{1}{c}{-0.006} & \multicolumn{1}{c}{0.003} & \multicolumn{1}{c}{-0.001} & \multicolumn{1}{c}{-0.001} & \multicolumn{1}{c}{0.011} & \multicolumn{1}{c}{0.009} \\
          & \multicolumn{1}{c}{(0.022)} & \multicolumn{1}{c}{(0.021)} & \multicolumn{1}{c}{(0.021)} & \multicolumn{1}{c}{(0.009)} & \multicolumn{1}{c}{(0.007)} & \multicolumn{1}{c}{(0.007)} & \multicolumn{1}{c}{(0.006)} & \multicolumn{1}{c}{(0.007)} & \multicolumn{1}{c}{(0.007)} & \multicolumn{1}{c}{(0.012)} & \multicolumn{1}{c}{(0.011)} \\
    Cuyo  & \multicolumn{1}{c}{0.028} & \multicolumn{1}{c}{0.013} & \multicolumn{1}{c}{0.010} & \multicolumn{1}{c}{0.012} & \multicolumn{1}{c}{0.008} & \multicolumn{1}{c}{0.007} & \multicolumn{1}{c}{0.009} & \multicolumn{1}{c}{0.009} & \multicolumn{1}{c}{0.009} & \multicolumn{1}{c}{0.014} & \multicolumn{1}{c}{0.019*} \\
          & \multicolumn{1}{c}{(0.018)} & \multicolumn{1}{c}{(0.017)} & \multicolumn{1}{c}{(0.016)} & \multicolumn{1}{c}{(0.008)} & \multicolumn{1}{c}{(0.008)} & \multicolumn{1}{c}{(0.008)} & \multicolumn{1}{c}{(0.006)} & \multicolumn{1}{c}{(0.007)} & \multicolumn{1}{c}{(0.007)} & \multicolumn{1}{c}{(0.010)} & \multicolumn{1}{c}{(0.010)} \\
    Pampeana & \multicolumn{1}{c}{0.025**} & \multicolumn{1}{c}{0.021} & \multicolumn{1}{c}{0.021} & \multicolumn{1}{c}{0.016**} & \multicolumn{1}{c}{0.016**} & \multicolumn{1}{c}{0.017**} & \multicolumn{1}{c}{0.010*} & \multicolumn{1}{c}{0.010} & \multicolumn{1}{c}{0.011} & \multicolumn{1}{c}{0.019**} & \multicolumn{1}{c}{0.018**} \\
          & \multicolumn{1}{c}{(0.012)} & \multicolumn{1}{c}{(0.013)} & \multicolumn{1}{c}{(0.013)} & \multicolumn{1}{c}{(0.006)} & \multicolumn{1}{c}{(0.007)} & \multicolumn{1}{c}{(0.007)} & \multicolumn{1}{c}{(0.006)} & \multicolumn{1}{c}{(0.007)} & \multicolumn{1}{c}{(0.006)} & \multicolumn{1}{c}{(0.008)} & \multicolumn{1}{c}{(0.008)} \\
    Patagonia & \multicolumn{1}{c}{0.039} & \multicolumn{1}{c}{0.022} & \multicolumn{1}{c}{0.027} & \multicolumn{1}{c}{0.026*} & \multicolumn{1}{c}{0.022} & \multicolumn{1}{c}{0.025*} & \multicolumn{1}{c}{0.017*} & \multicolumn{1}{c}{0.023*} & \multicolumn{1}{c}{0.023*} & \multicolumn{1}{c}{0.041**} & \multicolumn{1}{c}{0.029**} \\
          & \multicolumn{1}{c}{(0.026)} & \multicolumn{1}{c}{(0.024)} & \multicolumn{1}{c}{(0.023)} & \multicolumn{1}{c}{(0.015)} & \multicolumn{1}{c}{(0.014)} & \multicolumn{1}{c}{(0.014)} & \multicolumn{1}{c}{(0.010)} & \multicolumn{1}{c}{(0.013)} & \multicolumn{1}{c}{(0.012)} & \multicolumn{1}{c}{(0.017)} & \multicolumn{1}{c}{(0.013)} \\
    \multicolumn{5}{l}{Educational level (base = never attended) } &       &       &       &       &       &       &  \\
    Primary incomplete & \multicolumn{1}{c}{-0.035} & \multicolumn{1}{c}{-0.051} & \multicolumn{1}{c}{-0.052} & \multicolumn{1}{c}{-0.013} & \multicolumn{1}{c}{-0.026} & \multicolumn{1}{c}{-0.026} & \multicolumn{1}{c}{-0.010} & \multicolumn{1}{c}{-0.021} & \multicolumn{1}{c}{-0.023} & \multicolumn{1}{c}{-0.042} & \multicolumn{1}{c}{-0.041} \\
          & \multicolumn{1}{c}{(0.034)} & \multicolumn{1}{c}{(0.053)} & \multicolumn{1}{c}{(0.054)} & \multicolumn{1}{c}{(0.016)} & \multicolumn{1}{c}{(0.026)} & \multicolumn{1}{c}{(0.027)} & \multicolumn{1}{c}{(0.014)} & \multicolumn{1}{c}{(0.026)} & \multicolumn{1}{c}{(0.027)} & \multicolumn{1}{c}{(0.031)} & \multicolumn{1}{c}{(0.027)} \\
    Primary complete & \multicolumn{1}{c}{0.017} & \multicolumn{1}{c}{-0.025} & \multicolumn{1}{c}{-0.021} & \multicolumn{1}{c}{0.007} & \multicolumn{1}{c}{-0.014} & \multicolumn{1}{c}{-0.012} & \multicolumn{1}{c}{0.002} & \multicolumn{1}{c}{-0.006} & \multicolumn{1}{c}{-0.007} & \multicolumn{1}{c}{-0.018} & \multicolumn{1}{c}{-0.027} \\
          & \multicolumn{1}{c}{(0.047)} & \multicolumn{1}{c}{(0.062)} & \multicolumn{1}{c}{(0.063)} & \multicolumn{1}{c}{(0.025)} & \multicolumn{1}{c}{(0.035)} & \multicolumn{1}{c}{(0.035)} & \multicolumn{1}{c}{(0.017)} & \multicolumn{1}{c}{(0.023)} & \multicolumn{1}{c}{(0.024)} & \multicolumn{1}{c}{(0.025)} & \multicolumn{1}{c}{(0.024)} \\
    Secondary incomplete & \multicolumn{1}{c}{0.020} & \multicolumn{1}{c}{-0.038} & \multicolumn{1}{c}{-0.035} & \multicolumn{1}{c}{0.006} & \multicolumn{1}{c}{-0.018} & \multicolumn{1}{c}{-0.017} & \multicolumn{1}{c}{-0.002} & \multicolumn{1}{c}{-0.012} & \multicolumn{1}{c}{-0.014} & \multicolumn{1}{c}{-0.019} & \multicolumn{1}{c}{-0.027} \\
          & \multicolumn{1}{c}{(0.062)} & \multicolumn{1}{c}{(0.083)} & \multicolumn{1}{c}{(0.084)} & \multicolumn{1}{c}{(0.030)} & \multicolumn{1}{c}{(0.043)} & \multicolumn{1}{c}{(0.043)} & \multicolumn{1}{c}{(0.019)} & \multicolumn{1}{c}{(0.030)} & \multicolumn{1}{c}{(0.031)} & \multicolumn{1}{c}{(0.029)} & \multicolumn{1}{c}{(0.028)} \\
    Secondary complete & \multicolumn{1}{c}{-0.045} & \multicolumn{1}{c}{-0.129} & \multicolumn{1}{c}{-0.120} & \multicolumn{1}{c}{-0.036} & \multicolumn{1}{c}{-0.070} & \multicolumn{1}{c}{-0.065} & \multicolumn{1}{c}{-0.035} & \multicolumn{1}{c}{-0.051} & \multicolumn{1}{c}{-0.052} & \multicolumn{1}{c}{-0.036} & \multicolumn{1}{c}{-0.056} \\
          & \multicolumn{1}{c}{(0.082)} & \multicolumn{1}{c}{(0.117)} & \multicolumn{1}{c}{(0.116)} & \multicolumn{1}{c}{(0.037)} & \multicolumn{1}{c}{(0.054)} & \multicolumn{1}{c}{(0.054)} & \multicolumn{1}{c}{(0.026)} & \multicolumn{1}{c}{(0.040)} & \multicolumn{1}{c}{(0.041)} & \multicolumn{1}{c}{(0.037)} & \multicolumn{1}{c}{(0.039)} \\
    Superior incomplete & \multicolumn{1}{c}{-0.107} & \multicolumn{1}{c}{-0.194} & \multicolumn{1}{c}{-0.187} & \multicolumn{1}{c}{-0.089} & \multicolumn{1}{c}{-0.125*} & \multicolumn{1}{c}{-0.121*} & \multicolumn{1}{c}{-0.074*} & \multicolumn{1}{c}{-0.097*} & \multicolumn{1}{c}{-0.097*} & \multicolumn{1}{c}{-0.075} & \multicolumn{1}{c}{-0.092} \\
          & \multicolumn{1}{c}{(0.120)} & \multicolumn{1}{c}{(0.159)} & \multicolumn{1}{c}{(0.158)} & \multicolumn{1}{c}{(0.053)} & \multicolumn{1}{c}{(0.070)} & \multicolumn{1}{c}{(0.070)} & \multicolumn{1}{c}{(0.042)} & \multicolumn{1}{c}{(0.057)} & \multicolumn{1}{c}{(0.057)} & \multicolumn{1}{c}{(0.053)} & \multicolumn{1}{c}{(0.055)} \\
    Superior complete & \multicolumn{1}{c}{-0.104} & \multicolumn{1}{c}{-0.141} & \multicolumn{1}{c}{-0.133} & \multicolumn{1}{c}{-0.124**} & \multicolumn{1}{c}{-0.122} & \multicolumn{1}{c}{-0.117} & \multicolumn{1}{c}{-0.142**} & \multicolumn{1}{c}{-0.113*} & \multicolumn{1}{c}{-0.113*} & \multicolumn{1}{c}{-0.093} & \multicolumn{1}{c}{-0.113*} \\
          & \multicolumn{1}{c}{(0.137)} & \multicolumn{1}{c}{(0.167)} & \multicolumn{1}{c}{(0.167)} & \multicolumn{1}{c}{(0.061)} & \multicolumn{1}{c}{(0.072)} & \multicolumn{1}{c}{(0.072)} & \multicolumn{1}{c}{(0.063)} & \multicolumn{1}{c}{(0.067)} & \multicolumn{1}{c}{(0.067)} & \multicolumn{1}{c}{(0.063)} & \multicolumn{1}{c}{(0.066)} \\
    \multicolumn{5}{l}{Sector (base = Agricultural)} &       &       &       &       &       &       &  \\
    Construction & \multicolumn{1}{c}{1.666***} & \multicolumn{1}{c}{1.649***} & \multicolumn{1}{c}{1.643***} & \multicolumn{1}{c}{0.680***} & \multicolumn{1}{c}{0.704***} & \multicolumn{1}{c}{0.702***} & \multicolumn{1}{c}{0.193***} & \multicolumn{1}{c}{0.314***} & \multicolumn{1}{c}{0.310***} & \multicolumn{1}{c}{0.132*} & \multicolumn{1}{c}{0.143**} \\
          & \multicolumn{1}{c}{(0.338)} & \multicolumn{1}{c}{(0.361)} & \multicolumn{1}{c}{(0.364)} & \multicolumn{1}{c}{(0.122)} & \multicolumn{1}{c}{(0.121)} & \multicolumn{1}{c}{(0.123)} & \multicolumn{1}{c}{(0.063)} & \multicolumn{1}{c}{(0.091)} & \multicolumn{1}{c}{(0.092)} & \multicolumn{1}{c}{(0.066)} & \multicolumn{1}{c}{(0.060)} \\
    Electricity, gas, and water & \multicolumn{1}{c}{0.425} & \multicolumn{1}{c}{0.536} & \multicolumn{1}{c}{0.546} & \multicolumn{1}{c}{0.220} & \multicolumn{1}{c}{0.267*} & \multicolumn{1}{c}{0.272*} & \multicolumn{1}{c}{0.160} & \multicolumn{1}{c}{0.219*} & \multicolumn{1}{c}{0.223*} & \multicolumn{1}{c}{0.196*} & \multicolumn{1}{c}{0.172} \\
          & \multicolumn{1}{c}{(0.351)} & \multicolumn{1}{c}{(0.369)} & \multicolumn{1}{c}{(0.372)} & \multicolumn{1}{c}{(0.144)} & \multicolumn{1}{c}{(0.149)} & \multicolumn{1}{c}{(0.150)} & \multicolumn{1}{c}{(0.114)} & \multicolumn{1}{c}{(0.126)} & \multicolumn{1}{c}{(0.127)} & \multicolumn{1}{c}{(0.112)} & \multicolumn{1}{c}{(0.103)} \\
    Financial intermediation & \multicolumn{1}{c}{-0.087} & \multicolumn{1}{c}{0.000} & \multicolumn{1}{c}{0.015} & \multicolumn{1}{c}{-0.071} & \multicolumn{1}{c}{-0.044} & \multicolumn{1}{c}{-0.037} & \multicolumn{1}{c}{-0.130*} & \multicolumn{1}{c}{-0.090} & \multicolumn{1}{c}{-0.087} & \multicolumn{1}{c}{-0.107} & \multicolumn{1}{c}{-0.142*} \\
          & \multicolumn{1}{c}{(0.227)} & \multicolumn{1}{c}{(0.254)} & \multicolumn{1}{c}{(0.256)} & \multicolumn{1}{c}{(0.100)} & \multicolumn{1}{c}{(0.106)} & \multicolumn{1}{c}{(0.107)} & \multicolumn{1}{c}{(0.077)} & \multicolumn{1}{c}{(0.085)} & \multicolumn{1}{c}{(0.084)} & \multicolumn{1}{c}{(0.069)} & \multicolumn{1}{c}{(0.074)} \\
    Fishing & \multicolumn{1}{c}{0.152} & \multicolumn{1}{c}{0.381} & \multicolumn{1}{c}{0.394} & \multicolumn{1}{c}{-0.007} & \multicolumn{1}{c}{0.118} & \multicolumn{1}{c}{0.124} & \multicolumn{1}{c}{0.034} & \multicolumn{1}{c}{0.091} & \multicolumn{1}{c}{0.093} & \multicolumn{1}{c}{0.105} & \multicolumn{1}{c}{0.074} \\
          & \multicolumn{1}{c}{(0.280)} & \multicolumn{1}{c}{(0.299)} & \multicolumn{1}{c}{(0.301)} & \multicolumn{1}{c}{(0.145)} & \multicolumn{1}{c}{(0.167)} & \multicolumn{1}{c}{(0.168)} & \multicolumn{1}{c}{(0.105)} & \multicolumn{1}{c}{(0.137)} & \multicolumn{1}{c}{(0.136)} & \multicolumn{1}{c}{(0.129)} & \multicolumn{1}{c}{(0.129)} \\
    Food and accomm. & \multicolumn{1}{c}{-0.353} & \multicolumn{1}{c}{-0.350} & \multicolumn{1}{c}{-0.344} & \multicolumn{1}{c}{-0.088} & \multicolumn{1}{c}{-0.096} & \multicolumn{1}{c}{-0.093} & \multicolumn{1}{c}{-0.062} & \multicolumn{1}{c}{-0.069} & \multicolumn{1}{c}{-0.069} & \multicolumn{1}{c}{-0.059} & \multicolumn{1}{c}{-0.073} \\
          & \multicolumn{1}{c}{(0.243)} & \multicolumn{1}{c}{(0.260)} & \multicolumn{1}{c}{(0.262)} & \multicolumn{1}{c}{(0.103)} & \multicolumn{1}{c}{(0.107)} & \multicolumn{1}{c}{(0.108)} & \multicolumn{1}{c}{(0.068)} & \multicolumn{1}{c}{(0.078)} & \multicolumn{1}{c}{(0.078)} & \multicolumn{1}{c}{(0.055)} & \multicolumn{1}{c}{(0.051)} \\
    Human Health & \multicolumn{1}{c}{-0.377*} & \multicolumn{1}{c}{-0.275} & \multicolumn{1}{c}{-0.271} & \multicolumn{1}{c}{-0.049} & \multicolumn{1}{c}{-0.010} & \multicolumn{1}{c}{-0.008} & \multicolumn{1}{c}{-0.123} & \multicolumn{1}{c}{-0.066} & \multicolumn{1}{c}{-0.066} & \multicolumn{1}{c}{-0.081} & \multicolumn{1}{c}{-0.090} \\
          & \multicolumn{1}{c}{(0.222)} & \multicolumn{1}{c}{(0.246)} & \multicolumn{1}{c}{(0.249)} & \multicolumn{1}{c}{(0.101)} & \multicolumn{1}{c}{(0.104)} & \multicolumn{1}{c}{(0.105)} & \multicolumn{1}{c}{(0.078)} & \multicolumn{1}{c}{(0.083)} & \multicolumn{1}{c}{(0.083)} & \multicolumn{1}{c}{(0.066)} & \multicolumn{1}{c}{(0.062)} \\
    Industry & \multicolumn{1}{c}{0.465*} & \multicolumn{1}{c}{0.553*} & \multicolumn{1}{c}{0.561*} & \multicolumn{1}{c}{0.124} & \multicolumn{1}{c}{0.167} & \multicolumn{1}{c}{0.171} & \multicolumn{1}{c}{0.077} & \multicolumn{1}{c}{0.119} & \multicolumn{1}{c}{0.121} & \multicolumn{1}{c}{0.096} & \multicolumn{1}{c}{0.077} \\
          & \multicolumn{1}{c}{(0.271)} & \multicolumn{1}{c}{(0.294)} & \multicolumn{1}{c}{(0.297)} & \multicolumn{1}{c}{(0.149)} & \multicolumn{1}{c}{(0.155)} & \multicolumn{1}{c}{(0.157)} & \multicolumn{1}{c}{(0.093)} & \multicolumn{1}{c}{(0.114)} & \multicolumn{1}{c}{(0.113)} & \multicolumn{1}{c}{(0.093)} & \multicolumn{1}{c}{(0.088)} \\
    Mining & \multicolumn{1}{c}{0.584**} & \multicolumn{1}{c}{0.687**} & \multicolumn{1}{c}{0.700**} & \multicolumn{1}{c}{0.288**} & \multicolumn{1}{c}{0.319**} & \multicolumn{1}{c}{0.325**} & \multicolumn{1}{c}{0.233*} & \multicolumn{1}{c}{0.268**} & \multicolumn{1}{c}{0.272**} & \multicolumn{1}{c}{0.249**} & \multicolumn{1}{c}{0.219**} \\
          & \multicolumn{1}{c}{(0.268)} & \multicolumn{1}{c}{(0.284)} & \multicolumn{1}{c}{(0.286)} & \multicolumn{1}{c}{(0.138)} & \multicolumn{1}{c}{(0.139)} & \multicolumn{1}{c}{(0.140)} & \multicolumn{1}{c}{(0.117)} & \multicolumn{1}{c}{(0.118)} & \multicolumn{1}{c}{(0.117)} & \multicolumn{1}{c}{(0.105)} & \multicolumn{1}{c}{(0.101)} \\
    Other services & \multicolumn{1}{c}{-0.331} & \multicolumn{1}{c}{-0.301} & \multicolumn{1}{c}{-0.299} & \multicolumn{1}{c}{-0.053} & \multicolumn{1}{c}{-0.053} & \multicolumn{1}{c}{-0.052} & \multicolumn{1}{c}{0.059} & \multicolumn{1}{c}{0.098} & \multicolumn{1}{c}{0.092} & \multicolumn{1}{c}{0.011} & \multicolumn{1}{c}{0.005} \\
          & \multicolumn{1}{c}{(0.228)} & \multicolumn{1}{c}{(0.249)} & \multicolumn{1}{c}{(0.252)} & \multicolumn{1}{c}{(0.102)} & \multicolumn{1}{c}{(0.104)} & \multicolumn{1}{c}{(0.105)} & \multicolumn{1}{c}{(0.078)} & \multicolumn{1}{c}{(0.092)} & \multicolumn{1}{c}{(0.090)} & \multicolumn{1}{c}{(0.056)} & \multicolumn{1}{c}{(0.049)} \\
    Public adm. & \multicolumn{1}{c}{-0.166} & \multicolumn{1}{c}{-0.075} & \multicolumn{1}{c}{-0.068} & \multicolumn{1}{c}{-0.019} & \multicolumn{1}{c}{0.017} & \multicolumn{1}{c}{0.021} & \multicolumn{1}{c}{-0.084} & \multicolumn{1}{c}{-0.033} & \multicolumn{1}{c}{-0.034} & \multicolumn{1}{c}{-0.060} & \multicolumn{1}{c}{-0.076} \\
          & \multicolumn{1}{c}{(0.231)} & \multicolumn{1}{c}{(0.255)} & \multicolumn{1}{c}{(0.256)} & \multicolumn{1}{c}{(0.097)} & \multicolumn{1}{c}{(0.101)} & \multicolumn{1}{c}{(0.102)} & \multicolumn{1}{c}{(0.072)} & \multicolumn{1}{c}{(0.078)} & \multicolumn{1}{c}{(0.078)} & \multicolumn{1}{c}{(0.061)} & \multicolumn{1}{c}{(0.059)} \\
    Real estate and business & \multicolumn{1}{c}{-0.115} & \multicolumn{1}{c}{-0.114} & \multicolumn{1}{c}{-0.102} & \multicolumn{1}{c}{-0.027} & \multicolumn{1}{c}{-0.021} & \multicolumn{1}{c}{-0.015} & \multicolumn{1}{c}{-0.038} & \multicolumn{1}{c}{-0.038} & \multicolumn{1}{c}{-0.036} & \multicolumn{1}{c}{-0.057} & \multicolumn{1}{c}{-0.085} \\
          & \multicolumn{1}{c}{(0.216)} & \multicolumn{1}{c}{(0.269)} & \multicolumn{1}{c}{(0.270)} & \multicolumn{1}{c}{(0.099)} & \multicolumn{1}{c}{(0.107)} & \multicolumn{1}{c}{(0.107)} & \multicolumn{1}{c}{(0.079)} & \multicolumn{1}{c}{(0.085)} & \multicolumn{1}{c}{(0.084)} & \multicolumn{1}{c}{(0.060)} & \multicolumn{1}{c}{(0.061)} \\
    Teaching & \multicolumn{1}{c}{-0.296} & \multicolumn{1}{c}{-0.210} & \multicolumn{1}{c}{-0.199} & \multicolumn{1}{c}{-0.018} & \multicolumn{1}{c}{0.002} & \multicolumn{1}{c}{0.008} & \multicolumn{1}{c}{-0.103} & \multicolumn{1}{c}{-0.072} & \multicolumn{1}{c}{-0.073} & \multicolumn{1}{c}{-0.082} & \multicolumn{1}{c}{-0.110} \\
          & \multicolumn{1}{c}{(0.231)} & \multicolumn{1}{c}{(0.265)} & \multicolumn{1}{c}{(0.268)} & \multicolumn{1}{c}{(0.102)} & \multicolumn{1}{c}{(0.108)} & \multicolumn{1}{c}{(0.110)} & \multicolumn{1}{c}{(0.079)} & \multicolumn{1}{c}{(0.088)} & \multicolumn{1}{c}{(0.088)} & \multicolumn{1}{c}{(0.070)} & \multicolumn{1}{c}{(0.071)} \\
    Trade & \multicolumn{1}{c}{0.011} & \multicolumn{1}{c}{0.115} & \multicolumn{1}{c}{0.126} & \multicolumn{1}{c}{0.012} & \multicolumn{1}{c}{0.035} & \multicolumn{1}{c}{0.041} & \multicolumn{1}{c}{0.030} & \multicolumn{1}{c}{0.034} & \multicolumn{1}{c}{0.033} & \multicolumn{1}{c}{-0.003} & \multicolumn{1}{c}{-0.029} \\
          & \multicolumn{1}{c}{(0.306)} & \multicolumn{1}{c}{(0.325)} & \multicolumn{1}{c}{(0.328)} & \multicolumn{1}{c}{(0.139)} & \multicolumn{1}{c}{(0.139)} & \multicolumn{1}{c}{(0.141)} & \multicolumn{1}{c}{(0.075)} & \multicolumn{1}{c}{(0.086)} & \multicolumn{1}{c}{(0.086)} & \multicolumn{1}{c}{(0.061)} & \multicolumn{1}{c}{(0.055)} \\
    Transport and Comm. & \multicolumn{1}{c}{0.648**} & \multicolumn{1}{c}{0.695**} & \multicolumn{1}{c}{0.699**} & \multicolumn{1}{c}{0.383**} & \multicolumn{1}{c}{0.388**} & \multicolumn{1}{c}{0.390**} & \multicolumn{1}{c}{0.191**} & \multicolumn{1}{c}{0.271**} & \multicolumn{1}{c}{0.272**} & \multicolumn{1}{c}{0.172**} & \multicolumn{1}{c}{0.161**} \\
          & \multicolumn{1}{c}{(0.301)} & \multicolumn{1}{c}{(0.306)} & \multicolumn{1}{c}{(0.306)} & \multicolumn{1}{c}{(0.152)} & \multicolumn{1}{c}{(0.148)} & \multicolumn{1}{c}{(0.148)} & \multicolumn{1}{c}{(0.092)} & \multicolumn{1}{c}{(0.112)} & \multicolumn{1}{c}{(0.112)} & \multicolumn{1}{c}{(0.068)} & \multicolumn{1}{c}{(0.068)} \\
    \multicolumn{5}{l}{Firm Type (base = Public)} &       &       &       &       &       &       &  \\
    Private & \multicolumn{1}{c}{0.067} & \multicolumn{1}{c}{0.053*} & \multicolumn{1}{c}{0.050*} & \multicolumn{1}{c}{0.008} & \multicolumn{1}{c}{0.012} & \multicolumn{1}{c}{0.011} & \multicolumn{1}{c}{0.029} & \multicolumn{1}{c}{0.031*} & \multicolumn{1}{c}{0.019} & \multicolumn{1}{c}{0.016} & \multicolumn{1}{c}{0.021*} \\
          & \multicolumn{1}{c}{(0.043)} & \multicolumn{1}{c}{(0.028)} & \multicolumn{1}{c}{(0.029)} & \multicolumn{1}{c}{(0.020)} & \multicolumn{1}{c}{(0.013)} & \multicolumn{1}{c}{(0.013)} & \multicolumn{1}{c}{(0.022)} & \multicolumn{1}{c}{(0.017)} & \multicolumn{1}{c}{(0.014)} & \multicolumn{1}{c}{(0.010)} & \multicolumn{1}{c}{(0.012)} \\
    Other & \multicolumn{1}{c}{0.091} & \multicolumn{1}{c}{0.088} & \multicolumn{1}{c}{0.072} & \multicolumn{1}{c}{0.021} & \multicolumn{1}{c}{0.020} & \multicolumn{1}{c}{0.014} & \multicolumn{1}{c}{-0.017} & \multicolumn{1}{c}{-0.026} & \multicolumn{1}{c}{-0.035} & \multicolumn{1}{c}{-0.029*} & \multicolumn{1}{c}{0.004} \\
          & \multicolumn{1}{c}{(0.063)} & \multicolumn{1}{c}{(0.061)} & \multicolumn{1}{c}{(0.060)} & \multicolumn{1}{c}{(0.021)} & \multicolumn{1}{c}{(0.018)} & \multicolumn{1}{c}{(0.017)} & \multicolumn{1}{c}{(0.023)} & \multicolumn{1}{c}{(0.022)} & \multicolumn{1}{c}{(0.023)} & \multicolumn{1}{c}{(0.016)} & \multicolumn{1}{c}{(0.016)} \\
    \multicolumn{5}{l}{Firm Size (base = 5 or less employees)} &       &       &       &       &       &       &  \\
    between 6 and 40 & \multicolumn{1}{c}{-0.497**} & \multicolumn{1}{c}{0.029} & \multicolumn{1}{c}{0.042} & \multicolumn{1}{c}{-0.309**} & \multicolumn{1}{c}{0.023} & \multicolumn{1}{c}{-0.003} & \multicolumn{1}{c}{-0.300*} & \multicolumn{1}{c}{0.027*} & \multicolumn{1}{c}{0.280**} & \multicolumn{1}{c}{-0.004} & \multicolumn{1}{c}{0.043*} \\
          & \multicolumn{1}{c}{(0.225)} & \multicolumn{1}{c}{(0.039)} & \multicolumn{1}{c}{(0.041)} & \multicolumn{1}{c}{(0.148)} & \multicolumn{1}{c}{(0.014)} & \multicolumn{1}{c}{(0.022)} & \multicolumn{1}{c}{(0.163)} & \multicolumn{1}{c}{(0.014)} & \multicolumn{1}{c}{(0.121)} & \multicolumn{1}{c}{(0.010)} & \multicolumn{1}{c}{(0.021)} \\
    more than 40 & \multicolumn{1}{c}{-0.509**} & \multicolumn{1}{c}{0.029} & \multicolumn{1}{c}{0.052} & \multicolumn{1}{c}{-0.295**} & \multicolumn{1}{c}{0.027} & \multicolumn{1}{c}{0.007} & \multicolumn{1}{c}{-0.278*} & \multicolumn{1}{c}{0.033} & \multicolumn{1}{c}{0.284**} & \multicolumn{1}{c}{0.051*} & \multicolumn{1}{c}{0.074**} \\
          & \multicolumn{1}{c}{(0.220)} & \multicolumn{1}{c}{(0.046)} & \multicolumn{1}{c}{(0.049)} & \multicolumn{1}{c}{(0.145)} & \multicolumn{1}{c}{(0.017)} & \multicolumn{1}{c}{(0.025)} & \multicolumn{1}{c}{(0.160)} & \multicolumn{1}{c}{(0.020)} & \multicolumn{1}{c}{(0.119)} & \multicolumn{1}{c}{(0.028)} & \multicolumn{1}{c}{(0.027)} \\
    \multicolumn{5}{l}{Hourly labor income decile (base = 1st decile)} &       &       &       &       &       &       &  \\
    2nd decile & \multicolumn{1}{c}{0.012} & \multicolumn{1}{c}{-0.029} & \multicolumn{1}{c}{-0.022} & \multicolumn{1}{c}{0.015} & \multicolumn{1}{c}{-0.010} & \multicolumn{1}{c}{-0.007} & \multicolumn{1}{c}{-0.006} & \multicolumn{1}{c}{0.007} & \multicolumn{1}{c}{0.009} & \multicolumn{1}{c}{0.010} & \multicolumn{1}{c}{-0.006} \\
          & \multicolumn{1}{c}{(0.027)} & \multicolumn{1}{c}{(0.023)} & \multicolumn{1}{c}{(0.023)} & \multicolumn{1}{c}{(0.013)} & \multicolumn{1}{c}{(0.012)} & \multicolumn{1}{c}{(0.012)} & \multicolumn{1}{c}{(0.005)} & \multicolumn{1}{c}{(0.007)} & \multicolumn{1}{c}{(0.007)} & \multicolumn{1}{c}{(0.007)} & \multicolumn{1}{c}{(0.006)} \\
    3rd decile & \multicolumn{1}{c}{0.029} & \multicolumn{1}{c}{-0.024} & \multicolumn{1}{c}{-0.009} & \multicolumn{1}{c}{0.021} & \multicolumn{1}{c}{-0.015} & \multicolumn{1}{c}{-0.008} & \multicolumn{1}{c}{-0.007} & \multicolumn{1}{c}{0.011} & \multicolumn{1}{c}{0.013} & \multicolumn{1}{c}{0.035*} & \multicolumn{1}{c}{0.001} \\
          & \multicolumn{1}{c}{(0.035)} & \multicolumn{1}{c}{(0.028)} & \multicolumn{1}{c}{(0.027)} & \multicolumn{1}{c}{(0.016)} & \multicolumn{1}{c}{(0.014)} & \multicolumn{1}{c}{(0.013)} & \multicolumn{1}{c}{(0.008)} & \multicolumn{1}{c}{(0.011)} & \multicolumn{1}{c}{(0.011)} & \multicolumn{1}{c}{(0.018)} & \multicolumn{1}{c}{(0.010)} \\
    4th decile & \multicolumn{1}{c}{0.045} & \multicolumn{1}{c}{-0.015} & \multicolumn{1}{c}{0.008} & \multicolumn{1}{c}{0.025} & \multicolumn{1}{c}{-0.016} & \multicolumn{1}{c}{-0.004} & \multicolumn{1}{c}{-0.005} & \multicolumn{1}{c}{0.014} & \multicolumn{1}{c}{0.018*} & \multicolumn{1}{c}{0.059**} & \multicolumn{1}{c}{0.006} \\
          & \multicolumn{1}{c}{(0.040)} & \multicolumn{1}{c}{(0.036)} & \multicolumn{1}{c}{(0.033)} & \multicolumn{1}{c}{(0.020)} & \multicolumn{1}{c}{(0.019)} & \multicolumn{1}{c}{(0.017)} & \multicolumn{1}{c}{(0.007)} & \multicolumn{1}{c}{(0.011)} & \multicolumn{1}{c}{(0.011)} & \multicolumn{1}{c}{(0.028)} & \multicolumn{1}{c}{(0.010)} \\
    5th decile & \multicolumn{1}{c}{0.041} & \multicolumn{1}{c}{-0.015} & \multicolumn{1}{c}{0.013} & \multicolumn{1}{c}{0.019} & \multicolumn{1}{c}{-0.021} & \multicolumn{1}{c}{-0.007} & \multicolumn{1}{c}{-0.006} & \multicolumn{1}{c}{0.014} & \multicolumn{1}{c}{0.018*} & \multicolumn{1}{c}{0.077**} & \multicolumn{1}{c}{0.012} \\
          & \multicolumn{1}{c}{(0.038)} & \multicolumn{1}{c}{(0.035)} & \multicolumn{1}{c}{(0.032)} & \multicolumn{1}{c}{(0.019)} & \multicolumn{1}{c}{(0.020)} & \multicolumn{1}{c}{(0.017)} & \multicolumn{1}{c}{(0.009)} & \multicolumn{1}{c}{(0.010)} & \multicolumn{1}{c}{(0.009)} & \multicolumn{1}{c}{(0.036)} & \multicolumn{1}{c}{(0.014)} \\
    6th decile & \multicolumn{1}{c}{0.030} & \multicolumn{1}{c}{-0.036} & \multicolumn{1}{c}{-0.005} & \multicolumn{1}{c}{0.013} & \multicolumn{1}{c}{-0.031} & \multicolumn{1}{c}{-0.015} & \multicolumn{1}{c}{-0.014} & \multicolumn{1}{c}{0.005} & \multicolumn{1}{c}{0.010} & \multicolumn{1}{c}{0.081*} & \multicolumn{1}{c}{0.007} \\
          & \multicolumn{1}{c}{(0.042)} & \multicolumn{1}{c}{(0.044)} & \multicolumn{1}{c}{(0.037)} & \multicolumn{1}{c}{(0.020)} & \multicolumn{1}{c}{(0.022)} & \multicolumn{1}{c}{(0.018)} & \multicolumn{1}{c}{(0.010)} & \multicolumn{1}{c}{(0.010)} & \multicolumn{1}{c}{(0.009)} & \multicolumn{1}{c}{(0.041)} & \multicolumn{1}{c}{(0.016)} \\
    7th decile & \multicolumn{1}{c}{0.049} & \multicolumn{1}{c}{-0.019} & \multicolumn{1}{c}{0.014} & \multicolumn{1}{c}{0.015} & \multicolumn{1}{c}{-0.030} & \multicolumn{1}{c}{-0.014} & \multicolumn{1}{c}{-0.012} & \multicolumn{1}{c}{0.004} & \multicolumn{1}{c}{0.010} & \multicolumn{1}{c}{0.090**} & \multicolumn{1}{c}{0.013} \\
          & \multicolumn{1}{c}{(0.046)} & \multicolumn{1}{c}{(0.052)} & \multicolumn{1}{c}{(0.044)} & \multicolumn{1}{c}{(0.021)} & \multicolumn{1}{c}{(0.025)} & \multicolumn{1}{c}{(0.020)} & \multicolumn{1}{c}{(0.013)} & \multicolumn{1}{c}{(0.013)} & \multicolumn{1}{c}{(0.011)} & \multicolumn{1}{c}{(0.044)} & \multicolumn{1}{c}{(0.018)} \\
    8th decile & \multicolumn{1}{c}{0.056} & \multicolumn{1}{c}{-0.018} & \multicolumn{1}{c}{0.015} & \multicolumn{1}{c}{0.014} & \multicolumn{1}{c}{-0.032} & \multicolumn{1}{c}{-0.016} & \multicolumn{1}{c}{-0.012} & \multicolumn{1}{c}{0.001} & \multicolumn{1}{c}{0.007} & \multicolumn{1}{c}{0.087*} & \multicolumn{1}{c}{0.010} \\
          & \multicolumn{1}{c}{(0.050)} & \multicolumn{1}{c}{(0.059)} & \multicolumn{1}{c}{(0.050)} & \multicolumn{1}{c}{(0.024)} & \multicolumn{1}{c}{(0.028)} & \multicolumn{1}{c}{(0.024)} & \multicolumn{1}{c}{(0.017)} & \multicolumn{1}{c}{(0.017)} & \multicolumn{1}{c}{(0.015)} & \multicolumn{1}{c}{(0.044)} & \multicolumn{1}{c}{(0.021)} \\
    9th decile & \multicolumn{1}{c}{0.078} & \multicolumn{1}{c}{0.000} & \multicolumn{1}{c}{0.032} & \multicolumn{1}{c}{0.012} & \multicolumn{1}{c}{-0.033} & \multicolumn{1}{c}{-0.018} & \multicolumn{1}{c}{-0.014} & \multicolumn{1}{c}{-0.003} & \multicolumn{1}{c}{0.003} & \multicolumn{1}{c}{0.083*} & \multicolumn{1}{c}{0.008} \\
          & \multicolumn{1}{c}{(0.056)} & \multicolumn{1}{c}{(0.065)} & \multicolumn{1}{c}{(0.057)} & \multicolumn{1}{c}{(0.027)} & \multicolumn{1}{c}{(0.032)} & \multicolumn{1}{c}{(0.027)} & \multicolumn{1}{c}{(0.023)} & \multicolumn{1}{c}{(0.022)} & \multicolumn{1}{c}{(0.019)} & \multicolumn{1}{c}{(0.042)} & \multicolumn{1}{c}{(0.023)} \\
    10th decile & \multicolumn{1}{c}{0.142*} & \multicolumn{1}{c}{0.058} & \multicolumn{1}{c}{0.087} & \multicolumn{1}{c}{0.001} & \multicolumn{1}{c}{-0.042} & \multicolumn{1}{c}{-0.028} & \multicolumn{1}{c}{-0.024} & \multicolumn{1}{c}{-0.012} & \multicolumn{1}{c}{-0.007} & \multicolumn{1}{c}{0.068*} & \multicolumn{1}{c}{-0.000} \\
          & \multicolumn{1}{c}{(0.080)} & \multicolumn{1}{c}{(0.085)} & \multicolumn{1}{c}{(0.080)} & \multicolumn{1}{c}{(0.030)} & \multicolumn{1}{c}{(0.034)} & \multicolumn{1}{c}{(0.029)} & \multicolumn{1}{c}{(0.029)} & \multicolumn{1}{c}{(0.025)} & \multicolumn{1}{c}{(0.023)} & \multicolumn{1}{c}{(0.034)} & \multicolumn{1}{c}{(0.023)} \\
    =1 if Informal & \multicolumn{1}{c}{-0.509**} & \multicolumn{1}{c}{0.009} & \multicolumn{1}{c}{} & \multicolumn{1}{c}{-0.291*} & \multicolumn{1}{c}{0.038*} & \multicolumn{1}{c}{} & \multicolumn{1}{c}{-0.603***} & \multicolumn{1}{c}{-0.260**} & \multicolumn{1}{c}{} & \multicolumn{1}{c}{-0.103**} & \multicolumn{1}{c}{} \\
          (productive definition) & \multicolumn{1}{c}{(0.230)} & \multicolumn{1}{c}{(0.048)} & \multicolumn{1}{c}{} & \multicolumn{1}{c}{(0.154)} & \multicolumn{1}{c}{(0.021)} & \multicolumn{1}{c}{} & \multicolumn{1}{c}{(0.193)} & \multicolumn{1}{c}{(0.116)} & \multicolumn{1}{c}{} & \multicolumn{1}{c}{(0.045)} & \multicolumn{1}{c}{} \\
    =1 if Informal &       &       & \multicolumn{1}{c}{0.080**} & \multicolumn{1}{c}{} & \multicolumn{1}{c}{} & \multicolumn{1}{c}{0.040**} & \multicolumn{1}{c}{} & \multicolumn{1}{c}{} & \multicolumn{1}{c}{0.009} & \multicolumn{1}{c}{} & \multicolumn{1}{c}{-0.188**} \\
    (legal definition)  &       &       & \multicolumn{1}{c}{(0.038)} & \multicolumn{1}{c}{} & \multicolumn{1}{c}{} & \multicolumn{1}{c}{(0.018)} & \multicolumn{1}{c}{} & \multicolumn{1}{c}{} & \multicolumn{1}{c}{(0.014)} & \multicolumn{1}{c}{} & \multicolumn{1}{c}{(0.088)} \\
          &       &       &       & \multicolumn{1}{c}{} &       &       &       &       &       &       &  \\
    Observations & \multicolumn{1}{c}{383,865} & \multicolumn{1}{c}{289,424} & \multicolumn{1}{c}{289,424} & \multicolumn{1}{c}{383,865} & \multicolumn{1}{c}{289,424} & \multicolumn{1}{c}{289,424} & \multicolumn{1}{c}{383,865} & \multicolumn{1}{c}{289,424} & \multicolumn{1}{c}{289,424} & \multicolumn{1}{c}{289,424} & \multicolumn{1}{c}{289,424} \\
    R-squared & \multicolumn{1}{c}{0.523} & \multicolumn{1}{c}{0.484} & \multicolumn{1}{c}{0.485} & \multicolumn{1}{c}{0.470} & \multicolumn{1}{c}{0.444} & \multicolumn{1}{c}{0.446} & \multicolumn{1}{c}{0.321} & \multicolumn{1}{c}{0.298} & \multicolumn{1}{c}{0.295} & \multicolumn{1}{c}{0.193} & \multicolumn{1}{c}{0.240} \\
    \midrule
\end{longtable}
\end{ThreePartTable}

\end{landscape}

\end{document}